\documentclass[12pt]{article}
\usepackage{amsmath}
\usepackage{amssymb}
\usepackage{hyperref}
\usepackage{color}
\usepackage{graphicx}
\usepackage{float}

\newcommand{\dd}{\mathrm{d}}
\newcommand{\ep}{\epsilon} 
\newcommand{\te}{\tilde{\epsilon}}
\newcommand{\oo}{\theta} 
\newcommand{\bo}{\widehat{\theta}} 
\newcommand{\bpartial}{\overline{\partial}} 
  
\newcommand{\da}{\dot{\alpha}}  
\newcommand{\db}{\dot{\beta}}

\newcounter{Theorems}
\newcounter{Definitions}
\def\amklink#1#2{\href{http://andreimikhailov.com/#1}{\textcolor{blue}{\bf #2}}}

\begin{document}

\begin{titlepage}
\begin{flushright}

\end{flushright}

\begin{center}
{\Large\bf $ $ \\ $ $ \\
On worldsheet curvature coupling in pure spinor sigma-model
}\\
\bigskip\bigskip\bigskip
{\large Henrique Flores and Andrei Mikhailov${}^{\dag}$}
\\
\bigskip\bigskip
{\it Instituto de F\'{i}sica Te\'orica, Universidade Estadual Paulista\\
R. Dr. Bento Teobaldo Ferraz 271, 
Bloco II -- Barra Funda\\
CEP:01140-070 -- S\~{a}o Paulo, Brasil\\
}

\vskip 1cm
\end{center}

\begin{abstract}
   We discuss the relation between unintegrated and integrated vertex operators in
   string worldsheet theory, in the context of BV formalism. In particular, we
   clarify the origin of the Fradkin-Tseytlin term. We first consider the case of bosonic
   string, and then concentrate on the case of pure spinor superstring in $AdS_5\times S^5$.
   In particular, we compute the action of $b_0 - \bar{b}_0$ on the beta-deformation vertex.
   As a by-product, we formulate some new conjectures on general finite-dimensional vertices.
\end{abstract}

\vfill
{\renewcommand{\arraystretch}{0.8}%
\begin{tabular}{rl}
${}^\dag\!\!\!\!$ 
& 
\footnotesize{on leave from Institute for Theoretical and 
Experimental Physics,}
\\    
&
\footnotesize{ul. Bol. Cheremushkinskaya, 25, 
Moscow 117259, Russia}
\\
\end{tabular}
}

\end{titlepage}

\tableofcontents

\section{Introduction}

In a general curved background, the $b$-ghost of the pure spinor superstring
\cite{Berkovits:2000fe,Berkovits:2001ue} is not holomorphic:
\begin{equation}\label{BIsNotHolomorphic}
\bar{\partial}b  = Q(\ldots)
\end{equation}
On one hand, this is a problem, complicating the computation of scattering amplitudes.
On the other hand, this is a tip of an interesting mathematical structure. It was suggested
in \cite{Mikhailov:2016myt,Mikhailov:2016rkp} that in such cases the definition of the string measure should be modified, so that
the resulting measure should descend on the factorspace of metrics\footnote{or, more generally, of Lagrangian submanifolds of BV phase space} over diffeomorphisms.
The method of \cite{Mikhailov:2016rkp,Mikhailov:2016myt} is to first construct a pseudodifferential form equivariant with respect
to diffeomorphisms, and then obtain a base form using some connection.

This procedure can be also used to study the insertion of \underline{\bf un}integrated vertex operators.
Once we inserted unintegrated vertex operators, we should then integrate over the moduli space
of Riemann surfaces with marked points.
Let us first integrate, for each {\em fixed} complex structure on $\Sigma$, over the
positions of the marked points, postponing the integration over complex structure for later.
We interpret the result as the insertion of the \underline{\bf in}tegrated vertex operator.
It is usually assumed that to any \underline{\bf un}integrated vertex operator $V$ corresponds
some integrated
vertex operator $U$. The naive formula is:
\begin{equation}\label{NaiveRelation}
U = b_{-1}\overline{b}_{-1} V
\end{equation}
However, this naive formula does not always work correctly. First of all, in the pure spinor
formalism, $b$ is a rational function of the pure spinor fields. This, generally speaking,
leads to $U$ being a rational function of the pure spinors, with non-constant denominators.
It is not clear if such rational expressions should be allowed in the worldsheet action.
We will leave this question open. Instead, we discuss another issue: Eq. (\ref{NaiveRelation}) does not tell
us the whole truth about the curvature coupling (the Fradkin-Tseytlin term in the worldsheet action).
In this paper we will explain how to derive the
Fradkin-Tseytlin term in the action starting from the insertion of the
\underline{\bf un}integrated vertex operator $V$. We will construct, following the
prescription of \cite{Mikhailov:2016rkp,Mikhailov:2016myt}, the integration measure for integrating over the point of insertion of $V$.
We will show that the procedure of \cite{Mikhailov:2016rkp,Mikhailov:2016myt}
simplifies. This is mostly due to the existence of a relatively straightforward construction
of a connection on the space of Lagrangian submanifolds, as a principal bundle with the
structure group diffeomorphisms. The curvature of this connection is essentially equal
to the Riemann curvature of the worldsheet metric. The curvature term in the base form generates,
effectively, the dilaton coupling (the Fradkin-Tseytlin term) on the string worldsheet. Under certain
conditions, this reasoning leads (Section \ref{sec:GeneralTheoryOfVertexInsertions}) to the formula for the
deformation of the dilaton superfield:
\begin{equation}\label{PhiFromB0}
(b_0-\bar{b}_0)V = Q\Phi
\end{equation}
In general there are two contributions to $\Phi$: one from Eq. (\ref{NaiveRelation}) and another
from Eq. (\ref{PhiFromB0}).

\paragraph     {Eqs. (\ref{NaiveRelation}) and (\ref{PhiFromB0}) in the case of bosonic string}
In the case of bosonic string (Section \ref{sec:BosonicString}), the curvature coupling, generally speaking, comes from {\em both}
Eq. (\ref{NaiveRelation}) and Eq. (\ref{PhiFromB0}). The contribution from Eq. (\ref{NaiveRelation}) is due to
the fact that already the \underline{\bf un}integrated vertex operator contains the
curvature coupling: $c\bar{c}\sqrt{g}R\Phi$.

\paragraph     {Eqs. (\ref{NaiveRelation}) and (\ref{PhiFromB0}) in the case of pure spinor superstring}
In Section \ref{OPEofBwithBeta} we discuss  Eqs. (\ref{NaiveRelation}) and (\ref{PhiFromB0}) in the context
of the pure spinor superstring on $AdS_5\times S^5$.
In this case, the only source of the curvature coupling is Eq. (\ref{PhiFromB0}) --- the second line of Eq. (\ref{OmegaBase}).

The $b$-ghost is a rational function of the pure spinor (not a polynomial).
Therefore, the OPEs $b_{-1}\overline{b}_{-1} V$ and $(b_0-\bar{b}_0)V$ are also non-polynomial.
We explicitly evaluate $(b_0-\bar{b}_0)V$ in the particular case when $V$ is the beta-deformation vertex,
using  the $b_0$ and $\bar{b}_0$ from \cite{Berkovits:2010zz} --- see Section \ref{OPEofBwithBeta}.
At this time, we do not know any specific application of the formulas
of Section \ref{OPEofBwithBeta}. However, these computations inspired us to make some
conjectures about the unintegrated vertex operators --- see Sections \ref{sec:BetaDefAndGeneralizations} and \ref{sec:Conjectures}.

One interesting feature of the beta-deformation is the existence of non-physical vertex
operators\cite{Bedoya:2010qz,Mikhailov:2012id}. 
They normally cannot be put on a curved worldsheet, because of the anomaly. However, once we
allow denominators of the form $1\over \mbox{STr}(\lambda_L\lambda_R)$, it seems that there is
no obstacle, and the nonphysical vertices can be included. This at least means, that the
first few orders in the expansion in powers of $\varepsilon$ in Eq. (\ref{DeformationOfWorldsheetAction}) actually make sense in string perturbation theory.

\section{General theory of vertex insertions}\label{sec:GeneralTheoryOfVertexInsertions}
In this Section we will apply the prescription of \cite{Mikhailov:2016rkp,Mikhailov:2016myt} for the
vertex operators insertion.

\subsection{Use of BV formalism and notations}
In BV formalism, instead of integrating over the worldsheet complex
structures, we integrate over general families of Lagrangian submanifolds $L$ in BV phase space.
The space of all Lagrangian submanifolds is denoted $\rm LAG$.
\marginpar{$\rm LAG$}
In this paper, we will only consider a $6g-6$-dimensional subspace of $\rm LAG$,
which corresponds to variations of the complex structure.

We use the
\amklink{math/bv/notations/notations.html}{notations}
of \cite{Mikhailov:2016rkp}. The odd Poisson bracket will be denoted $\{\_,\_\}_{\rm BV}$, or just $\{\_,\_\}$.
\marginpar{$\{\_,\_\}$}
For a vector field $\xi$ on the BV phase space, generated by a BV Hamiltonian, we denote that
Hamiltonian $\underline{\xi}$:
\begin{equation}
\xi = \{\underline{\xi},\_\}_{\rm BV}
\end{equation}

\subsection{Use of worldsheet metric}
Classically, the string worldhseet action  depends on the worldsheet metric only through its
complex structure. 
Quantum mechanically, the computation of the path integral usually
involves the choice of the worldsheet {\em metric} (and not just complex structure), and then showing that in critical dimension the result
of the computation is actually Weyl-invariant ({\it i.e.} only depends on the complex structure).

In this paper, we will need a worldsheet metric also for another purpose: to define a connection
\footnote{there might be other choices of a connection, not requiring a metric} on the space of Lagrangian submanifolds as a principal bundle:
\begin{equation}\label{OverDiff}
{\rm LAG} \longrightarrow {{\rm LAG}\over {\rm Diff}}
\end{equation}
which we need to convert an equivariant form into a base form.
Suppose that we choose a metric for every complex structure. Then, we will explain
in Section \ref{sec:Connection}, this defines
a choice of horizontal directions, {\it i.e.} a connection on (\ref{OverDiff}) --- see
Eqs. (\ref{ConnectionTraceless}) and (\ref{ConnectionDeDonder}).

Given a complex structure, we will use the constant curvature metric of unit volume, which
always exists and is unique by the uniformization theorem \cite{donaldson2011riemann}.
(But other global choices of a metric would also be OK.)

\subsection{String measure}\label{sec:StringMeasure}

\paragraph     {Equivariant Master Equation}
String worldsheet theory, in the approach of \cite{Mikhailov:2016rkp,Mikhailov:2016myt}, comes with a PDF\footnote{pseudo-differential form} $\Omega^{\tt base}$ on LAG, which is base with respect to
$H = \mbox{Diff}$. It is obtained from the
\amklink{math/bv/omega/Equivariant_half-densities.html}{equivariant half-density}
$\rho^{\tt C}$, which satisfies
the equivariant Master Equation:
\begin{equation}\label{EquivariantMasterEquation}
   \Delta_{\rm can}\rho^{\tt C}(\xi) = \underline{\xi}\rho^{\tt C}(\xi)
\end{equation}
where $\xi\in {\bf h} =\mbox{Lie}(H)$ is the equivariant parameter, and $\underline{\xi}$ the
corresponding BV Hamiltonian.

\paragraph     {Expansion in powers of $\xi$}
Let us write $\rho^{\tt C}(\xi)$ as a product:
\begin{equation}
\rho^{\tt C}(\xi) = e^{a(\xi)}\rho_{1/2}
\end{equation}
where $\rho_{1/2}$ is a half-density satisfying the usual (not equivariant) Master Equation:
\begin{align}
  & \rho_{1/2} \;=\; \exp(S_{\rm BV})\quad \mbox{\tt\small ( $S_{\rm BV}$ is string worldsheet }
  \\  
  & \phantom{\rho_{1/2} \;=\; \exp(S_{\rm BV})\quad\quad} \mbox{\tt\small Master Action )}
  \nonumber\\  
  & \Delta_{\rm can}\rho_{1/2} \;=\; 0
    \label{RhoSatisfiesQME}
\end{align}
and $a(\xi)$ is a function on the BV phase space, $a(0)=0$.
For any function $f$ and half-density $\rho_{1/2}$, let us denote:
\begin{equation}
\Delta_{\rho_{1/2}} f = \rho_{1/2}^{-1}\Delta_{\rm can}(f \rho_{1/2}) - (-)^{\bar{f}} f \rho_{1/2}^{-1}\Delta_{\rm can}\rho_{1/2}
\end{equation}
Eqs. (\ref{EquivariantMasterEquation}) and (\ref{RhoSatisfiesQME}) imply:
\begin{equation}\label{CurvOfA}
   \Delta_{\rho_{1/2}}a(\xi) + {1\over 2}\left\{ a(\xi), a(\xi) \right\}_{\rm BV}
   \;=\; \underline{\xi}
\end{equation}

\subsubsection{$a(\xi)$ for bosonic string and for pure spinor string}
\paragraph     {For bosonic string} $a(\xi)$ is background-independent, linear in $\xi$, and given by a simple formula:
\begin{equation}
a(\xi)  = a^{(1)}\langle\xi\rangle = \int_{\Sigma}\xi^{\alpha}c^{\star}_{\alpha}
\end{equation}

\paragraph     {For pure spinor string} $a(\xi)$ is a complicated  background-dependent expression.
For background $AdS_5\times S^5$, the $a^{(1)}\langle\xi\rangle$
\amklink{math/pure-spinor-formalism/AdS5xS5/Action\_of\_diffeomorphisms.html}{was constructed} in \cite{Mikhailov:2017mdo},
where it was called $\Phi_{\xi}$. Schematically:
\begin{align}
  a^{(1)}\langle\xi\rangle \;=\; \int_{\Sigma}\;\;
  & (\xi\cdot\partial Z^M)A_{M\alpha}\lambda^{\star\alpha} 
    + (\xi\cdot\partial Z^M)B_M^NZ^{\star}_N +
  \\ 
  & +  (\partial Z^M)C_M^{\alpha}{\cal L}_{\xi}w_{\alpha}
    + D^{\alpha\beta}w_{\alpha}{\cal L}_{\xi}w_{\beta}
\end{align}
where:\\
\begin{tabular}{rl}
  --&\!\!\!$Z$ are coordinates on super-$AdS_5\times S^5$\\  
  --&\!\!\!$\lambda$ are pure spinors (both $\lambda_L$ and $\lambda_R$)\\  
  --&\!\!\!$A_{M\alpha}$, $B_M^N$, $C_M^{\alpha}$ and $D^{\alpha\beta}$ are some functions of $Z$,\\
    &\!\!\!and rational functions of pure spinors 
\end{tabular}

\subsubsection{Some assumptions}
BV formalism is
\amklink{math/bv/BV-formalism/Infinite_dimensional_case.html}{ill-defined in field-theoretic context}
, because $\Delta^{(0)}$ is ill-defined.
We will  assume that on local functionals $\Delta^{(0)}=0$. In other words, when $f_{\rm loc}$ is
a local functional of the string worldsheet fields:
\begin{equation}\label{DeltaZeroIsZero}
\Delta_{\rho_{1/2}} f_{\rm loc} = \{S_{\rm BV}, f_{\rm loc}\}
\end{equation}
We believe that it is possible justify this assumption in worldsheet perturbation theory,
but at this time our considerations are not rigorous.

\subsection{Equivariant unintegrated vertex}\label{sec:UnintegratedVertex}
\paragraph     {Stabilizer of a point}
Insertion of unintegrated vertex operator $V$ at a point on $p\in\Sigma$ leads to breaking of the
diffeomorphisms down to the subgroup $\mbox{St}(p)\subset \mbox{Diff}$ which preserves $p$.
Let ${\rm st}(p)$ denote the Lie algebra of $\mbox{St}(p)$:
\begin{align}
  \mbox{St}(p) \;=\;& \{g\in \mbox{Diff}\;|\; g(p)=p\}
  \\
  {\rm st}(p) \;=\;& \mbox{Lie}(\mbox{St}(p))
\end{align}
We will now explain how to construct an $\mbox{St}(p)$-equivariant form on  ${\rm LAG}$, and then
in Sections \ref{sec:Connection} and \ref{sec:IntegrationOfUnintegrated} how to construct a base form.

\paragraph     {Equivariantization of vertex}

Given an unintegrated vertex $V$, 
suppose that we can construct  for any $\xi_0\in {\rm st}(p)$ an equivariant vertex $V^{\tt C}(\xi_0)$,
satisfying\footnote{the subindex $\tt C$ stands for {\bf C}artan model of equivariant cohomology}:
\begin{equation}
V^{\tt C}(0) = V
\end{equation}
\begin{equation}\label{DeltaVisZero}
\Delta_{\rho^{\tt C}(\xi_0)}V^{\tt C}(\xi_0) = 0
\end{equation}
and:
\begin{equation}\label{VisEquivariant}
\{\underline{\xi_0},V^{\tt C}(\eta_0)\}_{\rm BV}  = \left.{d\over dt}\right|_{t=0} V^{\tt C}(e^{t[\xi_0\,,\,\_]}\eta_0)
\end{equation}
Under the conditions of Eqs. (\ref{DeltaVisZero}) and (\ref{VisEquivariant}) the product $V^{\tt C}(\xi_0)\rho^{\tt C}(\xi_0)$ defines an
${\rm st}(p)$-equivariant half-density satisfying the ${\rm st}(p)$-equivariant Master Equation:
\begin{equation}\label{DeltaPlusXiOnV}
   (\Delta_{\rm can} - \underline{\xi_0})\left(V^{\tt C}(\xi_0)\rho^{\tt C}(\xi_0) \right) = 0
\end{equation}
Any solution $V^{\tt C}(\xi_0)$ of Eq. (\ref{DeltaPlusXiOnV}) leads to ${\rm st}(p)$-equivariant pseudo-differential form:
\begin{equation}\label{EquivariantOmega}
   \Omega^{\tt C}(L,dL,\xi_0) = \int_{gL_0} \exp\left(
      \sigma\langle dL\rangle
   \right) \; V^{\tt C}(\xi_0)\rho^{\tt C}(\xi_0)
\end{equation}
Here $\sigma\langle dL\rangle$ is any BV Hamiltonian generating the infinitesimal
deformation $dL$ of $L$.

We can think of $V^{\tt C}(\xi_0)\rho^{\tt C}(\xi_0)$ as correction of the first order in $\epsilon$ to $\rho^{\tt C}(\xi_0)$ under
the deformation:
\begin{equation}
\rho\;\exp\left(a(\xi_0)\right) \rightarrow \rho\;\exp\left(a(\xi_0) + \varepsilon V^{\tt C}(\xi_0)\right)
\end{equation}
Eqs. (\ref{DeltaVisZero}) and (\ref{VisEquivariant}) imply:
\begin{align}
   & \left(\Delta_{\rho_{1/2}} + \{a(\xi_0),\_\}_{\rm BV}\right) V^{\tt C}(\xi_0) \;=\; 0
     \label{VisClosed} \\  
   & \{\underline{\xi_0},V^{\tt C}(\eta_0)\}_{\rm BV}  = \left.{d\over dt}\right|_{t=0} V^{\tt C}(e^{t[\xi_0\,,\,\_]}\eta_0)
   \label{VisInvariant}
\end{align}
The exact deformations, of the form:
\begin{align}\label{ExactDeformations}
  V^{\tt C}_{\tt exact}(\xi_0) \;=\;
  & \left(\Delta_{\rho_{1/2}} + \{a(\xi_0),\_\}_{\rm BV}\right) v^{\tt C}(\xi_0)
\end{align}
with $v^{\tt C}$ satisfying the equivariance condition $\{\underline{\xi_0},v^{\tt C}(\eta_0)\}_{\rm BV}  = \left.{d\over dt}\right|_{t=0} v^{\tt C}(e^{t[\xi_0\,,\,\_]}\eta_0)$
are considered trivial.

Consider the expansion of $V^{\tt C}(\xi_0)$ in powers of $\xi_0$:
\begin{equation}\label{ExpansionOfV}
V^{\tt C}(\xi_0) = V^{(0)} + V^{(1)}\langle \xi_0\rangle + V^{(2)}\langle \xi_0\otimes\xi_0\rangle + \ldots
\end{equation}
(We use angular brackets $\langle\ldots\rangle$ to highlight linearity, {\it i.e.} $f\langle x\rangle$ instead of $f(x)$
when $f$ is a linear functions of $x$.) In particular, Eq. (\ref{VisClosed}) implies at the linear order in $\xi$:
\begin{equation}\label{EqForV1}
\Delta_{\rho_{1/2}} V^{(1)}\langle\xi_0\rangle  + \{a^{(1)}\langle\xi_0\rangle, V^{(0)}\}_{\rm BV} = 0
\end{equation}
Equivariant vertex operators form a representation of the $D{\bf g}$ algebra discussed in \cite{Alekseev:2010gr}, 
the differential $d$ of \cite{Alekseev:2010gr} being represented by $\Delta_{\rho_{1/2}}$.

For our purpose, we will use a slightly different form of Eq. (\ref{EqForV1}). Let us return to Eq. (\ref{DeltaPlusXiOnV}).
At the linear order in $\xi_0$ it becomes:
\begin{equation}\label{EqForV1Alt}
   \Delta_{\rho_{1/2}} \left(
      a^{(1)}\langle \xi_0\rangle V^{(0)} + V^{(1)} \langle \xi_0\rangle
   \right) = \underline{\xi_0} V^{(0)}
\end{equation}
An exact $V$ corresponds to (see Eq. (\ref{ExactDeformations})):
\begin{align}
  V_{\rm exact}^{(0)} \;=\;
  & \Delta_{\rho_{1/2}} v^{(0)}
  \label{VExactAlt0}\\  
  V_{\rm exact}^{(1)}\langle \xi_0\rangle \;=\;
  & \Delta_{\rho_{1/2}} (a^{(1)}\langle \xi_0\rangle v^{(0)} + v^{(1)}\langle \xi_0\rangle)
    - \underline{\xi}^{(0)} v^{(0)}
    \label{VExactAlt1}
\end{align}

\vspace{10pt}\noindent
Eq. (\ref{EqForV1Alt}) is an equivalent form of Eq. (\ref{EqForV1}).
We will explain in Section \ref{sec:BosonicVsPure},
that in case of bosonic string it is more convenient to use Eq. (\ref{EqForV1}).
But in case of pure spinor string we use Eq. (\ref{EqForV1Alt}).

\subsection{A connection on $\Lambda\rightarrow \Lambda/St(p)$}\label{sec:Connection}
In order to integrate, we need to pass from equivariant $\Omega^{\tt C}$ to base $\Omega^{\tt base}$. This requires
a choice of a connection in the principal $St(p)$-bundle ${\rm LAG}\rightarrow {\rm LAG}/St(p)$. 
We will now define the connection by specifying the distribution
${\cal H}_0\subset TE|_S$ of horizontal vectors. We say that the vector belongs
to ${\cal H}_0$, if it is a linear combination of vectors of the following two classes:

\begin{itemize}
\item The {\bf first class} consists of the variations of the metric satisfying:
\begin{align}
  h^{\alpha\beta} \delta h_{\alpha\beta}  = \;& 0
                                                \label{ConnectionTraceless}
  \\   
  \nabla^{\alpha}\delta h_{\alpha\beta} = \;& 0
                                              \label{ConnectionDeDonder}
\end{align}
Such $\delta h_{\alpha\beta}$ can be identified as holomorphic or antiholomorphic quadratic differentials.
\item The {\bf second class} by definition consists of infinitesimal isometric (``rigid'')
   translations of  the disk $D_{\epsilon}$ of the small radius $\epsilon$. These are delta-function-like
   variations of the metric with the support on $\partial D_{\epsilon}$.
   They are always trivial in ${\rm LAG}/{\rm Diff}$, but nontrivial in ${\rm LAG}/St(p)$ when genus is greater
   than one.

   (This definition only works for the metric of constant negative curvature, because
   for generic metric $D_{\epsilon}$ does not have any infinitesimal isometries. In such cases,
   we can choose some lift to a vector field $v$ which is approximately isometry, in the
   sense that ${\cal L}_v g_{\alpha\beta} = O(|z|^2)$. Formulas do not change.)
\end{itemize}

\subsection{Base form and its integration}\label{sec:IntegrationOfUnintegrated}
Given a connection, we can
\amklink{math/bv/equivariant-cohomology/From_Cartan_To_Base.html}{construct a base form}
out of the equivariant form of Eq. (\ref{EquivariantOmega});
it is given by the following expression \cite{Mikhailov:2016rkp,Mikhailov:2016myt}:
\begin{equation}
   \Omega^{\tt base}(L,dL) =
   \int_{L} \exp({\sigma\langle dL|_{\rm hor}\rangle}) V^{\tt C}(F)\rho^{\tt C}(F)
\end{equation}
where $V^{\tt C}$ must satisfy Eqs. (\ref{DeltaVisZero}) and (\ref{VisEquivariant}), and $F$ is the curvature of our connection.
Here, as in Eq. (\ref{EquivariantOmega}), $\sigma\langle dL|_{\rm hor}\rangle$ is any BV Hamiltonian generating the
infinitesimal deformation, but we have to ``project'' the variation $dL$ to the horizontal subspace (using our
connection).

\vspace{10pt}\noindent
Let us consider the fiber bundle:
\begin{equation}
{{\rm MET}\over St(p)}\stackrel{\pi}{\longrightarrow} {{\rm MET}\over \mbox{Diff}}
\end{equation}
We want to integrate $\Omega$ over the cycle of the form
$\pi^{-1}c_{6g-6}$ where $c_{6g-6}$ is the fundamental cycle of the moduli space of Riemann
surfaces. Let us first integrate over the fiber (which is $\Sigma$). Our connection,
described in Section \ref{sec:Connection}, lifts the tangent vectors to the fiber  as
horizontal vectors of the second class, {\it i.e.}
as infinitesimal rigid translations of $D_{\epsilon}$. The curvature of our connection,
evaluated on a pair of vectors tangent to the fiber, takes values in
infinitesimal rigid rotations of $D_{\epsilon}$ and equals to the curvature of $\Sigma$.
Therefore $\Omega^{\tt base}$ is:
\begin{align}
  \Omega^{\tt base} \;=\;
  & \int e^S \Big[\;
    V^{(0)}
    \sigma\left\langle dL_{\rm hor} \right\rangle
    \wedge
    \sigma\left\langle dL_{\rm hor} \right\rangle
    \;+
  \nonumber\\
  &   \phantom{\int e^S\Big[}
    \;+\; V^{(0)}a^{(1)}\langle R\rangle
    + V^{(1)}\langle R\rangle
    \;\Big]
    \label{OmegaBase}
\end{align}
We will now explain this equation, first line first, and then the second.

\subsubsection     {First line of Eq. (\ref{OmegaBase})}
With our definition of the connection in Section \ref{sec:Connection}, the horizontal
projection $dL|_{\rm hor}$ is {\em an infinitesimal diffeomorphism}:
an infinitesimal  translation of the disk $D_{\epsilon}$ by $\left[\begin{array}{c}dz \cr d\bar{z}\end{array}\right]$.
Therefore, the corresponding BV Hamiltonian $\sigma\langle dL|_{\rm hor}\rangle$ is actually
$\Delta$-exact. Indeed, Eq. (\ref{CurvOfA}) implies that:
\begin{equation}\label{SigmaIsDelta}
   \sigma\langle dL|_{\rm hor}\rangle \;=\; \Delta_{\rho_{1/2}} a^{(1)}\langle u(dz,d\bar{z})\rangle
\end{equation}
Here $u(dz,d\bar{z})$ is the vector field on $\Sigma$ which is:\\
\begin{tabular}{rl}
  --&\!\!\!at the center of $D_{\epsilon}$ equals to $\left[\begin{array}{c}dz \cr d\bar{z}\end{array}\right]$ \\
  --&\!\!\!inside $D_{\epsilon}$ is an infinitesimal rigid translation\\
  --&\!\!\!outside of $D_{\epsilon}$ is zero\\
\end{tabular}\\
Since $a^{(1)}$ is a local functional on the string worldsheet,  Eqs. (\ref{SigmaIsDelta}) and (\ref{DeltaZeroIsZero})
imply:
\begin{equation}\label{SigmaVsPB}
\sigma\langle dL|_{\rm hor}\rangle \; =\; \{S_{\rm BV}\,,\,a^{(1)}\langle u(dz,d\bar{z})\rangle\}
\end{equation}

\paragraph     {Lemma-definition \arabic{Theorems}:\label{theorem:RestrictionOfDeltaPsi}}
For  any vector field $v$, the restriction of $\{S_{\rm BV}, a^{(1)}\langle v\rangle\}$ on $L$
is $\int b^{\alpha\beta} \nabla_{\alpha}v_{\beta}$:
\begin{equation}\label{RestrictionOfDeltaPsi}
   \left.\{S_{\rm BV},  a^{(1)}\langle v\rangle\}\right|_{L} = \int b^{\alpha\beta} \nabla_{\alpha}v_{\beta}
\end{equation}
We take Eq. (\ref{RestrictionOfDeltaPsi}) as the {\em definition} of $b^{\alpha\beta}$ (which is otherwise defined only
up to a $Q$-closed expression).
\paragraph     {Proof}   Let us consider the expansion of $S_{\rm BV}$ and the expansion of
${\cal V} = \{S_{\rm BV}, a^{(1)}\}$:
\begin{align}
  S_{\rm BV} \;=\; & S_0 + Q^A \phi^{\star}_A + \ldots
  \\
  \{S_{\rm BV}, a^{(1)}\} \;=\; & {\cal V}_0 + {\cal V}_1^A\phi^{\star}_A + \ldots
\end{align}
From $\{ S_{\rm BV}, \{ S_{\rm BV} , a^{(1)} \} \} = 0$ we derive:
\begin{align}
{\cal L}_Q {\cal V}_0 = {\cal L}_{{\cal V}_1} S_0 
\end{align}
Eq. (\ref{RestrictionOfDeltaPsi}) follows from the variation of $S_0$ under infinitesimal diffeomorphism being equal to
$\int T^{\alpha\beta}\nabla_{\alpha}v_{\beta}$, and from the vanishing of the off-shell cohomology in
ghost number $-1$ (we are working off-shell!).

Returning to Eq. (\ref{SigmaVsPB}),
Since $u$ is an isometry inside $D_{\epsilon}$ and zero outside $D_{\epsilon}$, we have:
\begin{equation}\label{Sa1OnL}
   \{S_{\rm BV}, a^{(1)}\langle u\rangle\}|_L 
   = \int_{\Sigma}\sqrt{g}\; b^{\alpha\beta}\nabla_{\alpha}u_{\beta}
   = \oint_{\partial D_{\epsilon}} dz^{\alpha} b_{\alpha\beta} u^{\beta}
\end{equation}
Therefore the first line in Eq. (\ref{OmegaBase}) contributes:
\begin{equation}
b_{-1}\bar{b}_{-1}V^{(0)}
\end{equation}

\subsubsection{Second line of Eq. (\ref{OmegaBase})}\label{sec:FirstLineOfOmegaBase}
Expressions like
$a^{(1)}\langle R\rangle$ and $V^{(1)}\langle R\rangle$ should be understood in the following way.
We think of the curvature $R$ as a two-form on the worldsheet with values in rotations of the tangent space:
\begin{equation}
R \in \Gamma\left(\Omega^2\Sigma \otimes so(T\Sigma)\right)
\end{equation}
In particular, if $\xi\in T_p\Sigma$ and $\eta\in T_p\Sigma$ are two tangent vectors, then
$R(\xi,\eta)$ at the point $p$  is an infinitesimal rotations of $T_p\Sigma$.
This infinitesimal rotation can be represented by a vector field $v$ with zero at the
point $p$. Let us ``truncate'' $v$ by putting it to zero outside $D_{\epsilon}$, {\it i.e.}
multiply $v$ by the function $\chi_{D_{\epsilon}}$ which is $1$ inside $D_{\epsilon}$ and $0$ outside.
By definition:
\begin{align}
  a^{(1)}\langle R(\xi,\eta)\rangle \stackrel{\tt def}{=}
  & a^{(1)}\langle \chi_{D_{\epsilon}}v \rangle
  \nonumber\\
  V^{(1)}\langle R(\xi,\eta)\rangle \stackrel{\tt def}{=}
  & V^{(1)}\langle \chi_{D_{\epsilon}}v \rangle
    \label{MeaningOfEvaluateOnR}
\end{align}
(This is an abbreviation, rather than a definition.)
In this context, Eq. (\ref{EqForV1Alt}) becomes:
\begin{equation}\label{a1V0plusV1}
   \Delta_{\rho_{1/2}}\left(
      V^{(0)} a^{(1)}\langle R\rangle 
      + V^{(1)}\langle R\rangle
   \right)\;=\; \{S_{\rm BV}, a^{(1)}\langle R\rangle\} V^{(0)}
\end{equation}

\paragraph     {In the case of pure spinor string} $\{a^{(1)}\langle R\rangle, V^{(0)}\} = 0$,
because $v$ in Eq. (\ref{MeaningOfEvaluateOnR}) is a vector field vanishing
at the point of insertion of $V^{(0)}$, and $V^{(0)}$ {\em does not contain derivatives}.
Therefore, the left hand side of Eq. (\ref{a1V0plusV1}) is
$\left\{
      S_{\rm BV}
      \,,\,
      a^{(1)}\langle R\rangle V^{(0)} + V^{(1)}\langle R\rangle
   \right\}$.
   When restricted to the Lagrangian submanifold, up to equations of motion\footnote{In spite of
     the fact that $\chi_{D_{\epsilon}}v$ of Eq. (\ref{MeaningOfEvaluateOnR}) is zero
     at the point of insertion of $V^{(0)}$, we
     cannot claim that $a^{(1)}\langle R\rangle|_L V^{(0)}|_L$ is zero.  This is because
     of the singularities in the OPE of the integrand of $a^{(1)}_L$ and $V^{(0)}$.
   }:
\begin{equation}\label{AltOnLag}
   Q\left(
      a^{(1)}\langle R\rangle|_L V^{(0)}|_L + V^{(1)}\langle R\rangle|_L
   \right) = \{S_{\rm BV}, a^{(1)}\langle R\rangle\}|_L V^{(0)}|_L
\end{equation}
We must stress that this equation is only valid under assumption
$\{a^{(1)}\langle R\rangle, V^{(0)}\} = 0$.
Generally speaking, instead of Eq. (\ref{AltOnLag}):
\begin{align}
  & Q\left(
      a^{(1)}\langle R\rangle|_L V^{(0)}|_L + V^{(1)}\langle R\rangle|_L
   \right) \;=
  \nonumber\\   
  =\; & \{S_{\rm BV}, a^{(1)}\langle R\rangle\}|_L V^{(0)}|_L
        - \left.\{a^{(1)}\langle R\rangle, V^{(0)}\}\right|_L
\end{align}
The computation of $\{S_{\rm BV}, a^{(1)}\langle R\rangle\}|_L$ uses Eq. (\ref{Sa1OnL}):
\begin{equation}
 \{S_{\rm BV}, a^{(1)}\langle R\rangle\}|_L V^{(0)}|_L = (b_0 - \bar{b}_0)V^{(0)}
\end{equation}
Therefore:
\begin{equation}
a^{(1)}\langle R\rangle|_L V^{(0)}|_L + V^{(1)}\langle R\rangle|_L = \sqrt{g}R\Phi
\end{equation}
where $\Phi$ satisfies:
\begin{equation}
Q\Phi = (b_0-\bar{b}_0)V^{(0)}
\end{equation}
To summarize, the total \underline{\bf in}tegrated vertex insertion corresponding
to the \underline{\bf un}integrated vertex $V^{(0)}$ is given by the expression:
\begin{align}
  & \int_{\Sigma}d^2 z\left( b_{-1}\bar{b}_{-1} V^{(0)} + \sqrt{g}R\Phi\right)
  \label{TotalIntegrated}\\
  &\mbox{\tt\small where $\Phi$  satisfies:}   Q\Phi = (b_0-\bar{b}_0)V^{(0)}
    \nonumber
\end{align}

\section{Brief review of the conventional description of the curvature coupling}
Here we will briefly review the ``standard'' derivation of the curvature coupling.

\vspace{10pt}
\noindent  Consider
the deformation of the worldsheet action by adding the {\em integrated} vertex operator:
\begin{equation}\label{DeformationOfWorldsheetAction}
S \mapsto S + \epsilon \int U
\end{equation}
where $\epsilon$ is a small ``deformation parameter''. Suppose that the deformed action
is classically BRST invariant. At the one loop level, we get:
\begin{align}
\partial^{\mu}j_{\mu}^{\rm BRST} \;=\; \alpha'(X + \sqrt{g}R Y)
\end{align}
where $X$ is a BRST-closed operator of conformal dimension $(1,1)$ and ghost number one,
and $Y$ is a BRST-closed expression of conformal dimension zero and ghost number
one\footnote{Notice that there is no $\sqrt{g} R V$ term in Eq. (\ref{DeformationOfWorldsheetAction}),
because there are no BRST-closed scalar operators $V$ of ghost number zero, other than
$1$ (the $1$ corresponding to the change in string coupling).}
In generic curved target-spaces, there is no BRST cohomology at ghost number 1 and conformal
dimension zero. Therefore, exists $\Phi$ such that:
\begin{equation}\label{PhiFromBRSTAnomaly}
Y = - Q_{\rm BRST}\Phi
\end{equation}
Also, there is no cohomology in conformal dimension $(1,1)$ and ghost number $1$, therefore
exists $U'$ such that $X = -Q U'$. These $U'$ and $\Phi$ can be absorbed into $U$:
\begin{equation}\label{AnomalyCancellation}
U\mapsto U + \alpha' U' + \alpha'\sqrt{g}R \Phi
\end{equation}
and the term $\Phi$ is the deformation of the dilaton.

\section{Bosonic string {\it vs} pure spinor string}\label{sec:BosonicVsPure}
\subsection{Main differences}
In {\bf pure spinor} string theory on $AdS_5\times S^5$:
\begin{itemize}
\item simplification: $V^{(0)}$ does not contain derivatives
\item complication: restriction of $a(\xi)$ on ``standard'' family of Lagrangian submanifolds is nonzero
\end{itemize}
In this case we need compute $\left.\left( a^{(1)}V^{(0)} + V^{(1)}\right)\right|_L$ (this is what deforms
the equivariant density), and we get it from Eq. (\ref{AltOnLag}) 

\vspace{10pt}\noindent
In {\bf bosonic} string theory:
\begin{itemize}
\item complication: $V^{(0)}$ contains at least derivatives of matter fields, and sometimes derivatives
   of ghosts
\item simplification: $a(\xi)$ is given by a simple formula: $a(\xi) = \xi^{\alpha}c^{\star}_{\alpha}$,
   and in particular its restriction to the standard Lagrangian submanifold is zero
\end{itemize}
In this situation we compute $V^{(1)}$ from Eq. (\ref{VisClosed}):
\begin{equation}
\{S_{\rm BV}, V^{(1)}\}  = - \{\xi^{\alpha}c^{\star}_{\alpha}\,,\,V^{(0)}\} = - \xi^{\alpha}{\partial\over\partial c^{\alpha}}V^{(0)}
\end{equation}
The exact vertex has:
\begin{align}
  V^{(0)}_{\rm exact} \;=\;
  & \{S_{\rm BV}, v^{(0)}\}
  \\   
  V^{(1)}_{\rm exact}\langle\xi\rangle \;=\;
  & \{S_{\rm BV}, v^{(1)}\langle\xi\rangle\} + \xi^{\alpha}{\partial\over\partial c^{\alpha}}v^{(0)}
  \\
  \ldots
\end{align}

\subsection{Bosonic string vertices as functions on BV phase space}\label{sec:BosonicString}
Consider bosonic string
on a general curved worldsheet. We work in BV formalism, our vertex operators are functions on the
\amklink{math/bv/bosonic-string/index.html}{BV phase space of bosonic string worldsheet}.

Let us start by considering the vertex corresponding to a ``gravitational wave'', {\it i.e.}
an infinitesimal deformation of the target space metric $G_{\mu\nu}$. We assume that
$G_{\mu\nu}$ satisfies transversality and linearized Einstein equations:
\begin{align}
  \partial^{\mu} G_{\mu\nu}=0
  \label{Transversality}
  \\
  \square G_{\mu\nu}= 0
  \label{Harmonicity}
\end{align}
(almost all gravitational waves can be obtained like this, except for some zero modes).
Let $h_{\alpha\beta}$ be the worldsheet metric, and $I^{\alpha}_{\beta}$ the corresponding complex
structure. We claim that the following vertex operator:
\begin{equation}\label{DefBosonicV0}
V^{(0)} = (Ic\cdot\partial X^{\mu})(c\cdot\partial X^{\mu})G_{\mu\nu}(x)
\end{equation}
satisfies:
\begin{equation}
\{S_{\rm BV}, V^{(0)}\}=0
\end{equation}
Let us prove this. The odd Poisson brackets with
\amklink{math/bv/bosonic-string/SolutionOfMasterEqn.html}{BV Master Action} are:
\begin{align}
  \{S_{\rm BV}, X\} \;=\;& {\cal L}_cX
  \\
  \{S_{\rm BV}, c\} \;=\;& {1\over 2}[c,c]
  \\
  \{S_{\rm BV}, I\} \;=\;& {\cal L}_c I
\end{align}
(Here ${\cal L}_c X$ is the same as $c\cdot \partial X$ --- the Lie derivative of $X$.)
\begin{align}
  & \{S_{\rm BV}, \;(Ic\cdot\partial X)(c\cdot\partial X)\} \;=\;
  \nonumber\\   
  =\;
  & (([{\cal L}_c, {\cal L}_{Ic}] - {\cal L}_{I[c,c]})X)\;{\cal L}_cX
    - ({\cal L}_{Ic} {\cal L}_cX)\;{\cal L}_cX + {1\over 2} ({\cal L}_{I[c,c]}X) \; {\cal L}_c X\;=
  \nonumber\\   
  =\;
  & ({\cal L}_c{\cal L}_{Ic}X) {\cal L}_c X - {1\over 2} ({\cal L}_{I[c,c]}X ) {\cal L}_cX
    \label{IntermediateForSBVV0}
\end{align}
Eq. (\ref{IntermediateForSBVV0}) follows from:
\begin{align}
  & {\cal L}_c{\cal L}_{Ic}X - {1\over 2} {\cal L}_{I[c,c]}X \;=\;
    {1\over 2}\iota_c^2\; d*d X
    \;=\;
    {1\over 2}\{S_{\rm BV}, \iota_c^2 X^{\star}\}
    \label{LcLIc}
\end{align}
In Eq. (\ref{LcLIc}) we identify $X^{\star}$ as a 2-form on the worldsheet, and contract
it two times with $c$. This operation can be characterized by saying that for every
local ({\it i.e.} given by a single integral over the worldsheet $\Sigma$) functional
$F[X]$:
\begin{equation}
   \left\{\iota_{\xi}\iota_{\eta} X^{\star}\,,\,F[X]\right\}\;=\;
   \iota_{\xi}\iota_{\eta} {\delta F\over\delta X}
\end{equation}
To prove Eq. (\ref{LcLIc}), let us choose the coordinates $(z,\bar{z})$ where
the complex structure is: $I{\partial\over\partial z} = i{\partial\over\partial z}$.
We denote $C=c^z$ and $\bar{C}=c^{\bar{z}}$,
{\it i.e.} $c\cdot\partial = C\partial + \bar{C}\bar{\partial}$ (with a slight abuse of
notations, we let $\partial$ denote also $\partial_z$). With these notations:
\begin{align}
  (C\partial + \bar{C}\bar{\partial})(iC\partial - i \bar{C}\bar{\partial})X
  -
  I(C\partial + \bar{C}\bar{\partial})^2 X\;=\;2i\bar{C}C \partial\bar{\partial}X
\end{align}
In order to actually insert $V^{(0)}$ we have to regularize it.
(Even when Eqs. (\ref{Transversality}) and (\ref{Harmonicity}) are satisfied,
we have the product of two $\partial X$ at the same point, which does not make sense without
regularization.)

\paragraph     {Regularization} We {\em regularize} $V^{(0)}$ by replacing every $X^{\mu}$
(including those acted on by $\partial$) with the averaged value:
\begin{equation}
  X(0,0)\;\mapsto\; {\cal N}_{\epsilon}\int d^2z \sqrt{g}
  \exp\left(-{1\over \epsilon}\mbox{dist}^2((z,\bar{z}),(0,0))\right)X(z,\bar{z})
 \label{OurRegularization}
\end{equation}
where $\mbox{dist}$ is the distance measured by the worldsheet metric,
$\epsilon \to 0$ the regularization parameter, 
and ${\cal N}_{\epsilon}$ is the normalization factor:
\begin{equation}
   {\cal N}_{\epsilon} = \left[
      \int d^2z \sqrt{g}  \exp\left(-{1\over \epsilon}\mbox{dist}^2((z,\bar{z}),(0,0))\right)
      \right]^{-1}
\end{equation}
When $c$ gets contracted with $\partial x$, we take the average of $c^{\alpha}\partial_{\alpha}x$.

\paragraph     {Renormalization} After specifying the regularization prescription, we have to
subtract infinities. 

\vspace{10pt}
\noindent Actually, with Eqs. (\ref{Transversality}) and (\ref{Harmonicity}) the subtraction
is not even needed, because the regularized $V^{(0)}$ remains finite when $\epsilon \to 0$.

\vspace{10pt}
\noindent But suppose that (having in mind extensions to string field theory) we want
to define our vertex  in a way which requires smooth extension off-shell, {\it i.e.} relaxing
of Eqs (\ref{Transversality}) and (\ref{Harmonicity}). Then, for our expression to remain
finite off-shell, we have to do a regularization. We define the {\em subtraction} as follows:
\begin{equation}\label{OurRenormalization}
  {\cal O}_{ren} = \exp\left(
  -\int d^2 z \int d^2 w \;
  \frac{\alpha'}{2} \ln \mathrm{dist}^{2}(z,\bar{z}; w,\bar{w})
  {\delta\over\delta X^{\mu}(z,\bar{z})}{\delta\over\delta X_{\mu}(w,\bar{w})}
  \right){\cal O}
\end{equation}
--- this removes the short distance singularity in $\langle X(z,\bar{z}) X(w,\bar{w})\rangle$.
Although this subtraction is diffeomorphism invariant, it is not Weyl invariant, and therefore
it does not commute with $\{S_{\rm BV},\_\}$.  The actual effect of the subtraction is:
\begin{equation}
   \mbox{lim}_{x\rightarrow y} \left(
      c^{\alpha}(x)c^{\beta}(y)
      {\partial\over\partial x^{\alpha}}{\partial\over\partial y^{\beta}}
      \log\mbox{dist}^2(x,y)
   \right) = {\alpha'\over 3}(c,Ic) R(x)
\end{equation}
This implies, that the unintegrated vertex annihilated by $\{S_{\rm BV},\_\}$ is:
\begin{align}
  & \left({\cal L}_cx^{\mu} {\cal L}_{Ic} x^{\nu} G_{\mu\nu}(x)\right)_{ren} +
  {\alpha'\over 3} (c, Ic) \Phi_{ren}R
  \\
  & \mbox{\tt\small where } \Phi =  G_{\mu}^{\mu} 
\end{align}
Therefore the curvature coupling arises from Eq. (\ref{NaiveRelation}), as $b_{-1}\bar{b}_{-1} \left( (c,Ic) R\; \Phi\right) = R\;\Phi$. (And this source of curvature coupling is not present in the
pure spinor case.)

\vspace{10pt}
\noindent
If we do not impose the condition (\ref{Transversality}), then Eq. (\ref{DefBosonicV0}) requires
modification. Additional terms should be added, such as {\it e.g.} $\mbox{div}\,c\,({\cal L}_c x^{\mu})A_{\mu}(x)$. With these extra terms, Eq. (\ref{PhiFromB0}) also contributes to the curvature
coupling.

\subsection{Ghost number one}
Cohomology at ghost number one is ({\it cp} Eq. (\ref{LcLIc})): 
\begin{align}
  W^{\mu} \;=\;
  & {\cal L}_{Ic} X^{\mu}  - {1\over 2}\iota^2_c X^{\mu\star}
  \\
  W^{\mu\nu} \;=\;
  & X^{[\mu} {\cal L}_{Ic} X^{\nu]} - {1\over 2}X^{[\mu} \iota^2_c X^{\nu]\star}
\end{align}
They are both already equivariant, because $\{a(\xi),W\}=0$, since $W$ does not
contain derivatives of $c$.
Notice that:
\begin{align}
  d W^{\mu} \;=\;
  & \left\{S_{\rm BV}\,,\, U^{\mu}\right\}
  \label{GhostNumberOneBosonicDescent}\\
  \mbox{\tt\small where }
  & U^{\mu} = *d X^{\mu} - \iota_c X^{\mu\star}
\end{align}
The proof of Eq. (\ref{GhostNumberOneBosonicDescent}) uses:
\begin{align}
  d {\cal L}_{Ic} X^{\mu} - {\cal L}_c *dX^{\mu} = \{S_{\rm BV},\iota_c X^{\mu\star}\}
\end{align}
As a consistency check, it should be true, at least in restriction to a reasonable Lagrangian
submanifold, that:
\begin{align}
  \iota_{\xi}U^{\mu} \;=\;
  & \left(\int_D \left\{S_{\rm BV},a^{(1)}\langle\xi\rangle\right\}\right)W^{\mu}
  \\
  \mbox{\tt\small where }
  & \left\{S_{\rm BV},a^{(1)}\langle\xi\rangle\right\} =
    ({\cal L}_{\xi}X^{\mu})X_{\mu}^{\star} + [\xi,c]c^{\star} + ({\cal L}_{\xi} g_{\alpha\beta})b^{\alpha\beta}
\end{align}
This is true on the
\amklink{math/bv/bosonic-string/LagrangianSubmanifold.html\#\%28part.\_section.Standard.Lagrangian.Submanifold\%29}{standard Lagrangian submanifold}
{\it i.e.} $c^{\star}=0$, $X^{\star}=0$. We did not explicitly check this for other Lagrangian
submanifolds. 

\subsection{Dilaton zero mode}
\paragraph     {Ghost dilaton}
Let us lift the expression $\partial c - \bar{\partial}\bar{c}$ of \cite{Belopolsky:1995vi} to the BV phase space as 
$v = \mbox{div}(Ic)$. The Cartan differential of $v$ is (see Eqs. (\ref{VExactAlt0}) and (\ref{VExactAlt1})):
\begin{align}
  V^{(0)} \;=\;
  & \{S_{\rm BV}, v\} = {\cal L}_c(\mbox{div}(Ic)) - {1\over 2}\mbox{div}(I[c,c])
    \label{GhostDilatonV0}
  \\     
  V^{(1)}\langle\xi_0\rangle \;=\;
  & \{a^{(1)}\langle\xi_0\rangle, v\} = \mbox{div}(I\xi_0)
    \label{GhostDilatonV1}
  \\
  V^{(\geq 2)}\langle\ldots\rangle \;=\;
  & 0
    \nonumber
\end{align}
The restriction of $V^{(0)}$ on the standard family is, on-shell,
$c\partial^2 c - \bar{c}\bar{\partial}^2 \bar{c}$.

\vspace{10pt}
\noindent
The base form corresponding to $V^{(1)}$ by the procedure of Section \ref{sec:IntegrationOfUnintegrated}
is $\sqrt{g}R$. Therefore, we should interpret $V^{(1)}$ as the unintegrated vertex operator
corresponding to the dilaton zero mode. However, $V^{(1)}$ by itself is not $\{S_{\rm BV},\_\}$-closed:
\begin{align}
  \{S_{\rm BV}, V^{(1)}\} = \mbox{tr}\Big(I\;
  & [{\cal L}_cI,{\cal L}_{\xi_0}I]\Big)\neq 0
  \\
  & \begin{array}{l}
     \mbox{\small\tt (commutator} \cr \mbox{\small\tt as matrices in $T_p\Sigma$)}
  \end{array}
\end{align}
\paragraph     {What is going on?}
The construction of the base form consists of the substitution of the curvature 2-form in place of $\xi_0$.
The way we construct connection in Section \ref{sec:Connection} it actually takes values in a smaller
subalgebra ${\rm st}(p, I_p)\subset {\rm st}(p)$, which consists of those vector fields which preserve
the complex structure {\em in the tangent space to the point $p$ of insertion}, {\it i.e.}
$I_p\in gl(T_p\Sigma)$. We observe that:
\begin{equation}
   \xi_0\in {\rm st}(p,I_p)\subset {\rm st}(p) \quad \Rightarrow\quad
   \{S_{\rm BV},V^{(1)}\langle\xi_0\rangle\} = 0
\end{equation}
(We must stress that, since $I$ is one of the BV fields, ${\rm st}(p,I_p)$ varies from point to point in the
BV phase space.)

\paragraph     {Equivalence of $V^{(0)}$ and $V^{(1)}$}
Eqs. (\ref{GhostDilatonV0})
and (\ref{GhostDilatonV1}) imply that the integrated vertex obtained from $V^{(1)}$
should be same as the one obtained from $V^{(0)}$. We can check this explicitly:
\begin{align}
  & \left(\oint dz^{\alpha}b_{\alpha\beta} \xi^{\beta}\right)
    \left(\oint dz^{\alpha}b_{\alpha\beta} \eta^{\beta}\right)
    V^{(0)}\;=\;
  \\    
  =\;& \left({\cal L}_{\xi}\mbox{div}(I\eta) - {1\over 2} \mbox{div}(I[\xi,\eta])\right) - (\xi\leftrightarrow\eta)\;=
  \\
  =\;& \mbox{div}(I[\xi,\eta]) = R(\xi,\eta)
\end{align}
We used the fact that, by the prescription of Section \ref{sec:Connection}, $\xi$ and $\eta$ are lifted
as isometries of a small neighborhood of the insertion point; in particular, the Lie derivative ${\cal L}_{\xi}$
commutes with the operations $I$ and $\mbox{div}$.

\subsection{Semirelative cohomology}
In our paper we identify the space of states as the cohomology of the {\em equivariant} complex,
as defined in Section \ref{sec:UnintegratedVertex}.

The usual definition is {\it via} the {\em semirelative} complex \cite{Belopolsky:1995vi}.
In the case of bosonic string, the cohomology is the same. Indeed, imposing the semirelative
condition $(b_0 - \bar{b}_0)V=0$ leads to two effects:\\   
\begin{tabular}{rl}
  {\bf Effect 1}&\!\!\!There are ghost number 2 cocycles, which should be thrown away\\  
                &because they are not annihilated by $b_0 - \bar{b}_0$. \\
                &Those are non-physical beta-deformations. \footnote{For pure spinor string, they are
  described in \cite{Mikhailov:2014qka} and references therein. The pure spinor case is similar}\\
  {\bf Effect 2}&\!\!\!The ghost-dilaton is $Q( \partial C - \bar{\partial}\bar{C} )$   --- would be BRST
                  exact \\
                &in the naive BRST complex, but $Q( \partial C - \bar{\partial}\bar{C} )$ is not\\  
                &annihilated by $b_0 -\bar{b}_0$. 
                  Therefore, the ghost-dilaton is actually\\
                &nontrivial
\end{tabular}\\
The equivariant complex gives the same result. For $V$ a nonphysical beta-deformation (Effect 1), 
$\{a(\xi),V\}$ is not just nonzero, but actually not even $\{S_{\rm BV},\_\}$-exact. Therefore, we cannot
``equivariantize'' such vertex in the sense of Section \ref{sec:UnintegratedVertex}. Therefore, such
states should be thrown away also in our approach. 

In case of Effect 2, we do admit $\partial C - \bar{\partial}\bar{C}$ (we present it as $\mbox{div}(Ic)$).
It is a perfectly valid cochain for us. However, our differential
is not just $Q_{BRST}$, or $\{S_{\rm BV},\_\}$. We actually have the equivariant differential,
which consists of two parts:
\begin{equation}
d_{\tt C}    =    \{S_{\rm BV},\_\}    +     \{ a\langle\xi\rangle ,  \_ \}
\end{equation}
$\left\{S_{\rm BV}, \partial C - \bar{\partial}\bar{C} \right\}$ is ghost dilaton, but the second term is also nonzero:
\begin{equation}
 \{ a\langle\xi\rangle , \partial C - \bar{\partial}\bar{C}  \}   =   \mbox{div} ( I \xi )
\end{equation}
Therefore, it is not the ghost-dilaton which is $d_{\tt C}$-exact,
but a sum of the ghost-dilaton and the expression $\mbox{div}(I\xi)$.
In other words, in our approach the ghost-dilaton is not $d$-exact,
but is $d$-equivalent to $\mbox{div}(I\xi)$.
Both expressions, when passing to the base form, result in $\sqrt{g}R$ --- the dilaton zero-mode.
This means that Effect 2 is also the same in our approach, as in the semirelative approach.

\section{Vertex operators of pure spinor superstring}\label{sec:PureSpinorString}
\subsection{Covariance of vertices}\label{sec:CovarianceOfVertices}
In this Section we will consider vertex operators of pure spinor superstring in $AdS_5\times S^5$.
We will restrict ourselves with only those vertex operators which transform in
\amklink{math/deformations-of-AdS/finite-dimensional-vertex/index.html}{finite-dimensional}
representations of $\bf g$ \cite{Mikhailov:2011af,Mikhailov:2017uoh}. We mainly consider the simplest example,
namely the beta-deformation, which
transforms in $({\bf g}\wedge {\bf g})_0\over {\bf g}$.  We also make some conjectures about
deformations transforming in other representations (``higher'' vertices, Section \ref{sec:Conjectures}).

Let $\cal H$ denote some subspace in the space of deformations, closed as a representation
of $\bf g$. We assume that the vertex is {\em covariant}.
This means that exists a map from $\cal H$ to space of vertices, commuting with the action
of $\bf g$. As was explained in \cite{Mikhailov:2011si}, under these conditions the all the
vertex operators in the given representation $\cal H$ are completely specified by a single
$\lambda$-dependent vector $v$ in the dual of $\cal H$:
\begin{equation}
v(\lambda_L, \lambda_R) \in {\cal H}'
\end{equation}
It should satisfy:
\begin{equation}
\rho(\lambda_L + \lambda_R) v = 0
\end{equation}
where $\rho(\lambda_L + \lambda_R)$ is the action of the element
$\lambda_L^{\alpha}t^3_{\alpha}  + \lambda_R^{\dot{\alpha}}t^1_{\dot{\alpha}}\in {\bf g}$ in ${\cal H}'$.
In this sense, the pure spinor BRST operator acts on ${\cal H}'$:
\begin{equation}\label{QOnv}
Q = \rho(\lambda_3 + \lambda_1) 
\end{equation}
In this Section we will study the case when $\cal H$ is finite-dimensional.
Then ${\cal H}' = {\cal H}$. We will consider those $\cal H$ which can be
constructed products of adjoint representations of $\bf g$, the simplest example
being the beta-deformation $({\bf g}\wedge {\bf g})_0\over {\bf g}$.
Such spaces are naturally related to the cochain complex of ${\bf g}$, which we will
now discuss.

\subsection{Lie algebra cohomology complex}\label{sec:LieAlgebraHomology}

Let us consider the Lie algebra cohomology complex of ${\bf g} = {\bf psu}(2,2|4)$ with
coefficients in a trivial representation. As a linear
space, it is the direct sum $\bigoplus\limits_{i=0}^{\infty} \Lambda^n{\bf g}'$, where ${\bf g}'$
is the dual space of $\bf g$. We use the fact that ${\bf g}$ has a supertrace, and identify
${\bf g}'$ with $\bf g$. The supertrace induces the pairing
\begin{equation}
\Lambda^n{\bf g}\otimes \Lambda^n{\bf g} \longrightarrow {\bf C}
\end{equation}
For example:
\begin{align}
  & \left\langle x\wedge y , z \wedge w \right\rangle \;=\;
  \\
  \;=\;
  &\mbox{STr}(yz) \mbox{STr}(xw)
   - (-1)^{\bar{x}\bar{y}} \mbox{STr}(xz) \mbox{STr}(yw)
\end{align}
The Lie superalgebra cohomology differential $d_{\rm Lie}$ acts as follows:
\begin{align}
  d_{\rm Lie} \;:\; & \Lambda^n{\bf g} \rightarrow \Lambda^{n+1}{\bf g}
  \\   
  & \left\langle d_{\rm Lie} x , y\wedge w\right\rangle \stackrel{{\rm def\;of\;}d_{\rm Lie}}{=} \left\langle x, [y,w] \right\rangle =
   \mbox{STr}(x[y,w])
\end{align}

\subsection{Vertex operators corresponding to global symmetries}\label{sec:GhostNumberOne}
The following element:
\begin{align}
\lambda_3 - \lambda_1  \in C^1{\bf g} = {\bf g}
\end{align}
is a nontrivial cocycle of $Q$. It corresponds to the unintegrated vertex operator:
\begin{equation}
V^{(0)}_a = \mbox{STr}(t_a g^{-1}(\lambda_3-\lambda_1)g)
\end{equation}

\subsection{Interplay between Lie algebra cohomology and pure spinor cohomology}
The $Q$-cocycle $\lambda_3-\lambda_1$ is not a $Q$-coboundary. However the Lie algebra differential
applied to it is a coboundary, if we allow denominator $1\over \mbox{STr}(\lambda_3\lambda_1)$:
\begin{align}\label{DLieOfGhostOneVertex}
  d_{\rm Lie} (\lambda_3 - \lambda_1) =
  Q \left(k^{\alpha\dot{\alpha}} t^3_{\alpha}\wedge ({\bf 1} - 2{\bf P}_{13})t^1_{\dot{\alpha}}\right)
\end{align}
The internal commutator of $k^{\alpha\dot{\alpha}}t^3_{\alpha}\wedge ({\bf 1} - 2{\bf P}_{13})t^1_{\dot{\alpha}}$ is nonzero, but is
$Q$-exact:
\begin{equation}
k^{\alpha\dot{\alpha}}\{t^3_{\alpha}, ({\bf 1} - 2{\bf P}_{13})t^1_{\dot{\alpha}}\} = {3\over 2}\{\lambda_3,\lambda_1\} = {3\over 4}Q(\lambda_3 + \lambda_1)
\end{equation}

\subsection{Beta-deformation and its generalizations}\label{sec:BetaDefAndGeneralizations}

\subsubsection{Definition}
The definition of the unintegrated vertex for beta-deformation given in \cite{Mikhailov:2011si,Bedoya:2010qz} is:
\begin{align}
  & V = B^{ab}W_aW_b
  \label{VertexBeta}\\
  & \mbox{\tt\small where }
    W_a = \mbox{STr}\left(t_ag^{-1}(\lambda_3-\lambda_1)g\right)
\end{align}
where $B^{ab}$ is a constant antisymmetric tensor, defined up to the equivalence relation:
\begin{equation}\label{EquivalenceRelationForB}
B^{ab}\simeq B^{ab} + f^{ab}{}_cA^c
\end{equation}
The beta-deformation transforms in the following the following representation of ${\bf psu}(2,2|4)$:
\begin{equation}\label{BetaSuperMultiplet}
{({\bf g}\wedge {\bf g})_0\over {\bf g}}
\end{equation}
where the factor over $\bf g$ accounts for the equivalence relation defined by the Eq. (\ref{EquivalenceRelationForB}).

This vertex operator defined in Eq. (\ref{VertexBeta}) is not strictly speaking covariant,
for the following reason. When we change $B^{ab}$ to $B^{ab} + f^{ab}{}_cA^c$, it changes by a BRST exact
expression:
\begin{align}
  V\longrightarrow & V + QW
  \\
  \mbox{\tt\small where }
  & W = \mbox{STr}\left(A g^{-1}(\lambda_3+\lambda_1)g\right)
\end{align}
It is possible to define the vertex which is strictly covariant: 
\begin{equation}
V' = V - \left\langle\, B\,,\, g^{-1}\left(
      \left[\Sigma,\lambda_3+\lambda_1\right]
      \wedge
      \left[\Sigma,\lambda_3+\lambda_1\right]
   \right)g\,
   \right\rangle
\end{equation}
where
\begin{equation}
\Sigma =\mbox{diag}(1,1,1,1,-1,-1,-1,-1)
\end{equation}
The difference between $V$ and $V'$ is a BRST-exact expression:
\begin{align}
  &
    \left\langle\, B\,,\, g^{-1}\left(
      \left[\Sigma,\lambda_3 + \lambda_1\right]
      \wedge
      \left[\Sigma,\lambda_3 + \lambda_1\right]
   \right)g\,
    \right\rangle\;=\; QX
  \label{XIsBRSTExact}\\  
  \mbox{\tt\small where }
  &
    X = - \left\langle\, B\,,\, g^{-1}\left(
      \Sigma
      \wedge
      \left[\Sigma,\lambda_3 + \lambda_1\right]
   \right)g\,
    \right\rangle
\end{align}
The definition of $X$ requires some work, because $\Sigma$ is not an element of ${\bf g} = {\bf psu}(2,2|4)$,
because $\mbox{STr}\Sigma\neq 0$.
Therefore, in order to define $X$, we need to lift $B$ from ${\bf g}\wedge {\bf g}$ to
${\bf su}(2,2|4)\wedge {\bf su}(2,2|4)$. There is no way to do it while preserving the ${\bf psu}(2,2|4)$-invariance. Therefore, $X$ does not transform as Eq. (\ref{BetaSuperMultiplet}). Still, Eq. (\ref{XIsBRSTExact}) holds, thus $V'$ {\em is}
BRST-equivalent to $V$.

\subsubsection{Alternative definition}
When $B$ satisfies the ``physicality'' condition $B^{ab}f_{ab}{}^c=0$, we can use the alternative
vertex:
\begin{align}\label{AlternativeBeta}
  \widetilde{V} = \mbox{STr}(\lambda_3\lambda_1)
  B^{ab}\Big\langle
  t_a\wedge t_b\;,\;
  g^{-1}\left(k^{\alpha\dot{\alpha}}t^3_{\alpha}\wedge {\bf P}_{13}t^1_{\dot{\alpha}}\right)g
  \Big\rangle
\end{align}
This alternative beta-deformation vertex is ``homogeneous'', in the sense that it has
a definite ghost number $(1,1)$. It is linear in $\lambda_3$ and in $\lambda_1$, because
the pre-factor $\mbox{STr}(\lambda_3\lambda_1)$ cancels the denominator in ${\bf P}_{13}$.

\paragraph     {Conjecture}
The vertex operator $\widetilde{V}$ defined by Eq. (\ref{AlternativeBeta}) is not BRST-exact.
If this is the case, then $\widetilde{V}$ is proportional to the beta-deformation vertex
of Eq. (\ref{VertexBeta}).
We leave the proof of this conjecture, and the computation of the proportionality coefficient,
for future work.

\subsection{Conjectures about higher finite-dimensional vertices}\label{sec:Conjectures}

\subsubsection{Recurrent construction of vertices}
Eq. (\ref{AlternativeBeta}) calls for generalization for higher finite-dimensional vertices
\cite{Mikhailov:2011af}. Let us consider the bicomplex:
\begin{equation}
d_{\rm tot} = Q + d_{\rm Lie}
\end{equation}
Eq. (\ref{DLieOfGhostOneVertex}) shows that:
\begin{align}
  Q v_2 =\;
  &  - d_{\rm Lie} v_1
  \\  
  \mbox{\tt\small where }
  & v_1 = \lambda_3 - \lambda_1
  \\
  & v_2 =  t^3_{\alpha}\wedge ({\bf 1} - 2{\bf P}_{13})t^1_{\dot{\alpha}}
\end{align}
Notice that the ghost number of $v_n$ is $2-n$.
\paragraph     {Conjecture:}
\begin{enumerate}
\item Exist $v_3,v_4,\ldots$ such that:
   \begin{equation}
      d_{\rm tot}\sum_{j=1}^{\infty} v_j = 0
   \end{equation}
\item For $j\geq 2$: $\left(\mbox{STr}(\lambda_3\lambda_1)\right)^jv_{2j}$ is a polynomial in
   $\lambda_3$ and $\lambda_1$, and is a covariant ghost number 2 vertex for the deformation
   corresponding to $\int d^4 x \mbox{tr} Z^{2 + j}$
\item For $j\geq 2$: $\left(\mbox{STr}(\lambda_3\lambda_1)\right)^{j+1}v_{2j + 1}$ is a polynomial in
   $\lambda_3$ and $\lambda_1$, and is a covariant ghost number 3 vertex, also corresponding
   to $\int d^4 x \mbox{tr} Z^{2 + j}$ as explained in \cite{Mikhailov:2014qka}.
\end{enumerate}
We leave the verification of these conjectures for future work.

\subsubsection{Infinitesimal deformations of worldsheet BV Master Action}
We will now describe another recurrent construction. As explained in \cite{Mikhailov:2017mdo}, the
pure spinor superstring in $AdS_5\times S^5$ is quasiisomorphic to the theory with the
following Master Action:
\begin{equation}\label{SBV}
S_{\rm BV} = \int \mbox{STr}\left( J_1\wedge ({\bf 1}- 2{\bf P}_{31}) J_3 \right)
\end{equation}
This is the integral over the worldsheet of the 2-form ${\cal B} = \mbox{STr}\left( J_1\wedge ({\bf 1}- 2{\bf P}_{31}) J_3 \right)$ which satisfies the property:
\begin{align}
  {\cal L}_Q{\cal B} =\;& d{\cal A}
  \\   
  \mbox{\tt\small where } & {\cal A} = \mbox{STr}(\lambda_3J_1 - \lambda_1J_3)
                            = \mbox{Str}\left((\lambda_3-\lambda_1)J\right)
\end{align}
It is natural to {\bf conjecture} that a vertex operator will correspond to an infinitesimal 
deformation of the action defined by Eq. (\ref{SBV}):
\begin{equation}
\Delta S_{\rm BV} = \int \left\langle\beta, J\wedge J\right\rangle
\end{equation}
Here $\beta$ is a rational function
of $\lambda$ with values in $\mbox{Hom}\left({\cal H}, {\bf g}\wedge {\bf g}\right)$,
where ${\cal H}$ is the space of deformations. 
The BRST invariance of the deformed action implies:
\begin{align}\label{QBeta}
  Q\beta = d_{\rm Lie}\alpha
\end{align}
Suppose that $\mbox{STr}(\lambda_3\lambda_1)\beta$ is a polynomial in $\lambda$. Then Eq. (\ref{QBeta}) implies that $\mbox{STr}(\lambda_3\lambda_1)\beta$ defines a $Q$-closed equivariant vertex
for ${\cal H}\otimes ({\bf g}\wedge {\bf g})_0$. We {\bf conjecture} that this vertex is
nontrivial ({\it i.e.} not BRST exact), although it may be BRST exact on a proper subspace $L\subset {\cal H}\otimes ({\bf g}\wedge {\bf g})_0$.
That means that, given a covariant vertex transforming in the representation ${\cal H}$, we
can build a new covariant vertex on the space of the larger spin representation
$\widetilde{\cal H} = {{\cal H}\otimes ({\bf g}\wedge {\bf g})_0\over L}$.
This gives a recurrent procedure for producing covariant vertices. We leave verification
of these conjectures for future work.

\section{OPE of $b$-ghost with beta-deformation vertex}\label{OPEofBwithBeta}
We will use the explicit formulas for the $b$-ghost collected in Section \ref{sec:BGhost}.

\subsection{General considerations}
At the leading order in $\alpha'$, we should have:
\begin{align}
  b_{zz} W_a \;=\; & {1\over z}(j_{az} + Q l_{az})
  \\
  b_{\bar{z}\bar{z}} W_a \;=\; & - {1\over \bar{z}}(j_{a\bar{z}} + Q l_{a\bar{z}})
\end{align}
where $l_{az}$, $l_{a\bar{z}}$ are some operators, and $j_{az} dz + j_{a\bar{z}} d\bar{z}$ is the global charge density; our definition of the charge density is such that:
\begin{equation}
   \left({1\over 2\pi i}\oint j_{az} dz + j_{a\bar{z}} d\bar{z} \right) W_b = f_{ab}{}^cW_c
\end{equation}
Notice:
\begin{align}
  j_{za} W_b \;=\;& {1\over 2 z} f_{ab}{}^c W_c + \ldots
  \\
  j_{\bar{z}a} W_b \;=\;& - {1\over 2\bar{z}} f_{ab}{}^c W_c + \ldots
\end{align}
(where $\ldots$ can include $\log z$ but not $z^{-1}$)
Therefore:
\begin{align}
  (b_0 - \bar{b}_0) V \;=\;
  & \oint\left( dz z b_{zz} - d\bar{z}\bar{z}b_{\bar{z}\bar{z}} \right) V\;=
  \nonumber\\
  \;=\;
  & B^{ab}f_{ab}{}^cW_c + Q\left[B^{ab}\left(\oint l_a\right) W_b\right]
    \label{GeneralFormulaForB0V}
\end{align}
One is tempted to say that Eq. (\ref{GeneralFormulaForB0V}) implies that $V$ is annihilated
by $b_0 - \bar{b}_0$, in cohomology, once $B$ satisfies the physicality condition
$B^{ab} f_{ab}{}^c=0$. 
However, notice that the expression $f_{ab}{}^cW_c$ is anyway $Q$-exact (and even
${\bf g}$-covariantly $Q$-exact)
since we allow denominator $1\over \mbox{STr}(\lambda_3\lambda_1)$, see Section \ref{sec:BRST-triviality}.

\subsection{Explicit computation}

The operator $(b_0-\bar{b}_0)V$ is a sum of two terms: the term with the ghost number $(1,0)$
and the term with the ghost number $(0,1)$. The term with the ghost number $(0,1)$ is:
\begin{align}
&\frac{2}{\mathrm{STr} (\lambda_3\lambda_1)}
\Bigg( \{ [\lambda_{1} , t^2_{-m}],\lambda_{3} \} \wedge [ t^2_m,\lambda_{1} ]
\, - \,  \{[\lambda_3 , t^2_{-m}],\lambda_1\}\wedge [t^2_m,\lambda_1] \Bigg) 
\nonumber \\
\nonumber \\
&- \kappa^{\beta\dot{\beta}}\{t^3_{\beta}, \lambda_1\}\wedge t^1_{\dot{\beta}}
\end{align}
and the term with the ghost number $(1,0)$ is equal, with the minus sign, to the same
expression with $\lambda_3\leftrightarrow\lambda_1$ and exchanged dotted and undotted indices.
Transform:
\begin{align}
  & - {2 \over \mbox{STr}(\lambda_3\lambda_1)} \{[\lambda_3 , t^2_{-m}],\lambda_1\}\wedge [t^2_m,\lambda_1]
    \;
\nonumber  \\
\nonumber  \\
  =\;
  & - {2\kappa^{\dot{\beta}\beta}\over \mbox{STr}(\lambda_3\lambda_1)}
    \{[\lambda_3,\overline{\{\lambda_1,t^1_{\dot{\beta}}\}}_{\rm STL}],\lambda_1\}\wedge t^3_{\beta}\;
\nonumber \\
\nonumber \\
  \;=\;
  & {2\kappa^{\dot{\beta}\beta}\over \mbox{STr}(\lambda_1\lambda_3)}
    \{[\lambda_3,\overline{\{\lambda_1,t^1_{\dot{\beta}}\}}_{\rm STL}],\lambda_1\}\wedge t^3_{\beta}\;
\nonumber \\
\nonumber \\
  \;=\;
  & - {2 \kappa^{\dot{\beta}\beta}\over \mbox{STr}(\lambda_1\lambda_3)}
    \{[\overline{\{\lambda_1,t^1_{\dot{\beta}}\}}_{\rm STL},\lambda_3],\lambda_1\}\wedge t^3_{\beta}\;
\nonumber \\
\nonumber \\
  \;=\;
  & \kappa^{\dot{\beta}\beta}\{t^1_{\dot{\beta}},\lambda_1\}\wedge t^3_{\beta}
\end{align}
where we used the explicit form of the pure spinor projector $\mathrm{P}_{31}$ that can be
found in \cite{Mikhailov:2017mdo}.
Thus we arrive at:
\begin{align}
  {2\over \mbox{STr}(\lambda_3\lambda_1)} \{[\lambda_1 , t^2_{-m}],\lambda_3\}\wedge [t^2_m,\lambda_1]
  - Q_R\left( \kappa^{\beta\dot{\beta}}t_{\beta}^3\wedge t^1_{\dot{\beta}}\right)
\end{align}
Adding the ``mirror'' term with the ghost number $(1,0)$, we arrive at:
\begin{align}
  & (b_0 - \bar{b}_0) \left\langle
    B^{ab}g(t_a\wedge t_b)g^{-1}\,,\,(\lambda_3-\lambda_1)\wedge (\lambda_3-\lambda_1)
    \right\rangle \;=\;
    Q\Phi
  \\  
  \mbox{\tt\small where: }
  \nonumber\\     \Phi\;=\;
  & \Big\langle B^{ab}g(t_a\wedge t_b)g^{-1}\,,\, 
    \label{OurDilaton}\\  
  & \phantom{\Big\langle}
    {2[t_{-m}^2,\lambda_1]\wedge [t_m^2,\lambda_1] + 2[t_{-m}^2,\lambda_3]\wedge [t_m^2,\lambda_3]
    \over \mbox{Str}(\lambda_3\lambda_1)}
    - \kappa^{\beta\dot{\beta}}t^3_{\beta}\wedge t^1_{\dot{\beta}}
    \Big\rangle
    \nonumber
\end{align}
Up to $Q$-exact terms, we can also take:
\begin{align}
  \Phi\;=\;
  & \Big\langle B^{ab}g(t_a\wedge t_b)g^{-1}\,,\, 
    \label{OurDilaton1}\\  
  & \phantom{\Big\langle}
    {-4[t_{-m}^2,\lambda_3]\wedge [t_m^2,\lambda_1] + 2[\{\lambda_3,\lambda_1\},t_{-m}^2]\wedge t_m^2
    \over \mbox{Str}(\lambda_3\lambda_1)}
    - \kappa^{\beta\dot{\beta}}t^3_{\beta}\wedge t^1_{\dot{\beta}}
    \Big\rangle
    \nonumber
\end{align}

\subsection{Discussion}
In this Section we will compare our proposed
Eq. (\ref{TotalIntegrated}):
\begin{align}\label{TotalIntegratedAgain}
  \int U \;=\;
  & \int_{\Sigma}d^2 z\left( b_{-1}\bar{b}_{-1} V^{(0)} + \sqrt{g}R\Phi\right)
  \\   
  & \mbox{\tt\small where $\Phi$ is given by Eq. (\ref{OurDilaton}) }
\end{align}
with the standard approach to the beta-deformation \cite{Bedoya:2010qz}. The most obvious observation is that
the ``dilaton superfield'' $\Phi$ of
Eq. (\ref{OurDilaton}) contains pure spinors (while the ``standard'' dilaton superfield,
obviously, does not). Therefore, they are certainly not the same. We will now explain that
there are two reasons for the difference.

\paragraph     {\underline{First reason:} $b_{-1}\bar{b}_{-1} V^{(0)}$ is different from the standard
  integrated vertex on flat worldsheet.}
The standard integrated vertex on flat worldsheet is\cite{Mikhailov:2011si,Bedoya:2010qz}:
\begin{equation}
B^{ab}j_a\wedge j_b
\end{equation}
In our approach here, it is the $b_{-1}\bar{b}_{-1} V^{(0)}$ of Eq. (\ref{TotalIntegratedAgain}).
This is {\em not} equal to $B^{ab}j_a\wedge j_b$, but differs from it by a $Q$-exact expression,
which we have not explicitly computed\footnote{since we have not explicitly computed $b_{-1}\bar{b}_{-1} V^{(0)}$}:
\begin{equation}\label{StandardVsBB}
B^{ab}j_a\wedge j_b = dz\wedge d\bar{z} b_{-1}\bar{b}_{-1} V^{(0)} + QX
\end{equation}
Notice that the BRST operator is only nilpotent on-shell:
\begin{equation}
Q^2 = {\partial S\over\partial w_1}{\partial\over\partial w_3} + (1\leftrightarrow 3)
\end{equation}
Therefore, the $QX$ on the RHS of Eq. (\ref{StandardVsBB}) deforms the BRST operator:
\begin{equation}
   Q\mapsto Q +
   \left(
      {\partial X\over\partial w_1}{\partial\over\partial w_3} + (1\leftrightarrow 3)
   \right)
\end{equation}
This leads to the change in the BRST anomaly, and, by the mechanism of Eqs. (\ref{PhiFromBRSTAnomaly}), (\ref{AnomalyCancellation}),
to the change of the Fradkin-Tseytlin term.

When we modify the unintegrated vertex:
\begin{equation}
V^{(0)} \mapsto \widetilde{V}^{(0)} = V^{(0)} + Q W^{(0)}
\end{equation}
The change in $\Phi$, {\it i.e.} $\widetilde{\Phi} - \Phi$, should satisfy:
\begin{equation}
Q (\widetilde{\Phi} - \Phi) = (b_0 - \bar{b}_0)QW^{(0)}
\end{equation}
Under the assumption that $(L_0 - \bar{L}_0)W^{(0)} =0$ this can be solved by taking:
\begin{equation}\label{ChangeOfDilatonUnderBRSTTransformation}
\widetilde{\Phi} = \Phi - (b_0 - \bar{b}_0)W^{(0)}
\end{equation}
Suppose that we were able to find such $W^{(0)}$ that $\widetilde{V}^{(0)}$ is polynomial in pure spinors.
Then, the curvature coupling also changes, according to Eq. (\ref{ChangeOfDilatonUnderBRSTTransformation}),

\paragraph     {\underline{Second reason:} we have not required the vanishing of $B^{ab}f_{ab}{}^c$.}
In fact, $\Phi$ of Eq. (\ref{OurDilaton1}) can be presented as:
\begin{align}
  \Phi\;=\;
  & B^{ab}\left(X_{[ab]} + \Big\langle g(t_a\wedge t_b)g^{-1}\,,\, 
    {2[\{\lambda_3,\lambda_1\},t_{-m}^2]\wedge t_m^2
    \over \mbox{Str}(\lambda_3\lambda_1)}
    \Big\rangle\right)
\end{align}
where $X_{[ab]}$ is defined in Eq. (\ref{Xab}). Since $QX_{[ab]}$ is proportional to $f_{ab}{}^c$,
the term $B^{ab}X_{[ab]}$ can be dropped when $B$ has zero internal commutator, {\it i.e.} $B^{ab}f_{ab}{}^c=0$. In that case, we have just:
\begin{equation}
\Phi\;=\;
 B^{ab}\Big\langle g(t_a\wedge t_b)g^{-1}\,,\, 
    {2[\{\lambda_3,\lambda_1\},t_{-m}^2]\wedge t_m^2
    \over \mbox{Str}(\lambda_3\lambda_1)}
    \Big\rangle
\end{equation}
We see that imposing the condition $B^{ab}f_{ab}{}^c=0$ ``considerably simplifies'' the expression
for the dilaton superfield. But still the resulting expression is a rational function of
$\lambda$'s.

\section*{Acknowledgements}
We would like to thank Nathan Berkovits, Ana L\'{u}cia Retore and Matheus Lize for useful discussions.
The work of A.M. was partially supported by RFBR 18-01-00460.

\appendix

\section{Technical details}
\subsection{MATHEMATICA code}
MATHEMATICA code for computations in $AdS_5\times S^5$ sigma-model is
\href{https://github.com/Henriquemfl/Pure-spinor-in-AdS5-Mathematica/tree/master}{\textcolor{blue}{\bf available on GitHub}}.

\subsection{Conventions and notations for $AdS_{5} \times S^{5}$ string}

We begin introducing some notation that will be useful through out the calculation.
Our notation is largely based on references \cite{Mikhailov:2007mr, Bedoya:2010qz}.

\paragraph{Constant Grassmann parameters}
The target space is a supermanifold, a coset of the Lie supergroup $PSU(2,2|4)$.
As usual \cite{Shvarts:1985fe}, treating the supermanifold, we introduce a ``pool'' of constant Grassmann
parameters $\ep, \ep', \ep'',\ldots$. We can construct the ``$\ep, \ep', \ep'',\ldots$-points'' of the supermanifold
$PSU(2,2|4)$ as formal expressions of the form, for example $\exp\left(\epsilon \mu^{\alpha}t^3_{\alpha} + \epsilon' \mu^{\dot{\alpha}}t^1_{\dot{\alpha}}\right)$
where $\mu^{\alpha}$ and $\mu^{\dot{\alpha}}$ are some spinors with real number components. In addition to these constant
Grassmann parameters, there are string worldsheet fields $\theta_L^{\alpha}$ and $\theta_R^{\dot{\alpha}}$; therefore we also have:
$\exp\left(\theta_L^{\alpha}t^3_{\alpha}\right)$ --- another element of the supergroup.

\paragraph{Superconformal generators and Casimir conventions}
An element in the superconformal algebra $\mathbf{g} = \mathrm{psu}(2,2|4)$ 
will be represented according to its $\mathbf{Z}_{4}$ grading,

\begin{align}
&t = t^{0}_{[mn]} \oplus t^{1}_{\dot{\alpha}} \oplus t^{2}_{m} \oplus t^{3}_{\alpha} \nonumber \\
\texttt{where} \nonumber \\
&\, t^{0}_{[mn]} \in \mathbf{g}_{0}, \quad t^{1}_{\dot{\alpha}} \in \mathbf{g}_{1}, \quad t^{2}_{m} \in \mathbf{g}_{2} \quad \text{and} \quad t^{3}_{\alpha} \in \mathbf{g}_{3}
\end{align}

\noindent
Latin letters are vector indices and greek letters are spinor indices. The bosonic
generators are boosts and rotations, given by $t^{0}_{[mn]}$, and translations denoted $t^{2}_{m}$. 
The fermionic generators are the right supersymmetries, $t^{1}_{\dot{\alpha}}$, and the left supersymmetries,
$t^{3}_{\alpha}$, with both spinors in the $d=10$ Majorana-Weyl representation. The vector space $\mathbf{g}_{2}$ is
the sum of the tangent vector spaces of $\mathrm{AdS}_{5}$ and $S^{5}$; $m\in\{0,\ldots,9\}$.

For a finite-dimensional representation, the invariant bilinear form is given by the
supertrace:

\begin{equation}
\mathrm{str} \left( t_{m}^{2} t_{n}^{2} \right) = \kappa_{mn}, \quad \mathrm{str} \left( t^{3}_{\alpha} t^{1}_{\dot{\alpha}} \right) = \kappa_{\alpha \dot{\alpha}}
\quad \texttt{and} \quad \mathrm{str} \left( t^{1}_{\dot{\alpha}} t^{3}_{\alpha} \right) = \kappa_{\dot{\alpha} \alpha}
\end{equation}

\noindent
where $\kappa_{\alpha \dot{\alpha}}$ and $\kappa_{mn}$ are Casimir tensors.

\subsection{The $b$-ghost} \label{sec:BGhost}
The $b$-ghost satisfies:
\begin{align}
  Q_{L} b_{zz} \;=\;& T_{zz},
  \\
  Q_{R} b_{zz} \;=\;& 0
\end{align}
where $T_{zz}$ is the holomorphic stress-energy tensor. The $\bar{b}_{\bar{z}\bar{z}}$ is defined by the same formula
with $Q_L$ exchanged with $Q_{R}$ and $T_{zz}$ replaced with $T_{\bar{z}\bar{z}}$.
The solutions of these equations are given by\cite{Berkovits:2008ga,Berkovits:2010zz}:
\begin{equation}\label{BInAdS}
b_{zz} = - \frac{ \mathrm{str} \left( \lambda_{1} \left[ J_{2z} \Sigma, J_{1z} \right] \right) }{ \mathrm{str} \left( \lambda_{3} \lambda_{1} \right) } 
+ \frac{1}{2} \mathrm{str} \left( \mathrm{P}_{13} \omega_{1z} J_{3z} \right)
\end{equation}
and
\begin{equation}\label{BBarInAdS}
\overline{b}_{\bar{z} \bar{z}} = + \frac{ \mathrm{str} \left( \lambda_{3} \left[ J_{2 \bar{z}} \Sigma, J_{3 \bar{z}} \right] \right) }{ \mathrm{str} \left( \lambda_{3} \lambda_{1} \right)}
+ \frac{1}{2} \mathrm{str} \left( \mathrm{P}_{31} \omega_{3\bar{z}} J_{1 \, \bar{z}} \right)
\end{equation}
where $\mathrm{P}_{13}$ and $\mathrm{P}_{31}$ are
\amklink{math/pure-spinor-formalism/AdS5xS5/Subspaces\_in\_superconformal\_algebra.html\#\%28part.\_.Tangent\_and\_normal\_space\_to\_pure\_spinor\_cones\%29}{some projectors}
.
These projectors are
needed because the pure spinor momenta $\omega_{1z}$ and $\omega_{3\bar{z}}$ are
defined up to gauge transformations of the form:
\begin{equation}
\delta_{u} \omega_{3z} = \left[ u_{z}, \lambda_{1} \right], \, \texttt{ and } \,
\delta_{u} \omega_{1\bar{z}} = \left[ u_{\bar{z}}, \lambda_{3} \right], 
\quad \end{equation}

\noindent
for both $u_{z}$ and $u_{\bar{z}}$ in $\mathbf{g}_{2}$.
Therefore, the projectors are constructed to satisfy 

\begin{equation}
\mathrm{P}_{13} \delta_{u} \omega_{1\bar{z}} = 0
\quad \texttt{ and } \quad
\mathrm{P}_{31} \delta_{u} \omega_{3z} = 0.
\end{equation}
Explicit formulas for $\mathrm{P}_{13}$ and $\mathrm{P}_{31}$ as rational
functions of the pure spinor variables can be found in \cite{Mikhailov:2017mdo}.

\vspace{10pt}

\noindent It is an open question to prove that the expressions
$\mathrm{str} \left( \mathrm{P}_{13} \omega_{1z} J_{3z} \right)$ and
$\mathrm{str} \left( \mathrm{P}_{31} \omega_{3\bar{z}} J_{1 \, \bar{z}} \right)$
are well-defined in the quantum theory.

\vspace{10pt}

\noindent
Lemma \ref{theorem:RestrictionOfDeltaPsi} implies that  $b$ given by Eqs. (\ref{BInAdS}) and (\ref{BBarInAdS}) coincides with $\Delta\Psi|_L$ up to a
$Q$-closed expression. We have not verified this explicitly.

\paragraph{Parametrization of $\mathrm{AdS}_{5} \times S^{5}$}
We will work with the conventions of \cite{Mikhailov:2007mr}.
The coordinates in $\mathrm{AdS}_{5} \times S^{5}$ are given by $\left(x, \oo, \bo \right)$ 
such that

\begin{equation}
x= x^{m}(z, \bar{z}) t_{m}^{2}  , \quad \oo = \oo^{\alpha} (z, \bar{z})  t_{\alpha}^{3} , \quad \bo = \bo^{\dot{\alpha}} (z, \bar{z}) t_{\dot{\alpha}}^{1}.
\end{equation}

\noindent
Each of these coordinates 
lifts to an element in $PSU(2,2|4)$ given by
\begin{equation}
g(x, \oo, \bo) = \exp \left( \frac{1}{R} \oo + \frac{1}{R} \bo \right) \exp \left( \frac{1}{R} x \right)
\label{coordinates}
\end{equation}
where $R$ is the AdS radius.

\paragraph{The pure spinor action}
The $AdS_{5} \times S^{5}$ pure spinor string action is

\begin{equation}
S = \frac{R^{2}}{ \pi} \int \dd^{2} z \,\, \mathrm{str} \biggl( \frac{1}{2} J_{2z} J_{2\bar{z}} + \frac{3}{4} J_{1z} J_{3\bar{z}} + \frac{1}{4} J_{3z} J_{1\bar{z}} + 
\omega_{1z} D_{\bar{z}} \lambda_{3} + \omega_{3z} D_{\bar{z}} \lambda_{1} + N_{0z} N_{0\bar{z}} \biggr)
\label{purespinoraction}
\end{equation}

\noindent
with the covariant derivatives defined as
\begin{subequations}
\begin{equation}
   D_{\bar{z}} \lambda_{3} = \partial_{\bar{z}} \lambda_{3} + [J_{0\bar{z}} , \lambda_{3} ],
   \quad
   D_{z} \lambda_{1} = \partial_{z} \lambda_{1} + [J_{0z} , \lambda_{1} ]
\end{equation}
and the Lorentz currents for the ghosts given by

\begin{equation}
N_{0z} = - \left\{ \omega_{1z} , \lambda_{1} \right\}, \quad  N_{0\bar{z}} = - \left\{ \omega_{3\bar{z}}, \lambda_{3} \right\}.
\end{equation}
\end{subequations}

\noindent
The pure spinor action is built out of the right-invariant currents:

\begin{equation}
J = - \dd g \, g^{-1} = - \partial_{z} g \, g^{-1} \dd z - \partial_{\bar{z}} g\, g^{-1} \dd \overline{z},
\end{equation}

\noindent
where $g$ is given by Eq. (\ref{coordinates}). These currents decompose according to the conformal weight
and the $\mathbf{Z}_{4}$ grading. We write $J = J_{0} + J_{1} + J_{2} + J_{3}$ to highlight the grading structure,
and we observe that under local Lorentz symmetry $J_{0}$ transforms as a connection
while $J_{1}$, $J_{2}$ and $J_{3}$ transform in the adjoint representation.

\paragraph{OPE between $b$-ghost and global vertex}
With these definitions, the OPE between the $b$-ghost and unintegrated global symmetry
becomes to $1$-loop order:

\begin{align}
\biggl \langle \ep \left( b_{0} - \overline{b}_{0} \right) V[\tilde{\ep}] (0) e^{-S_{i}} \biggr \rangle &=  
\biggl \langle \biggl( \oint \frac{ \dd z}{2 \pi i} \,\, z  \ep b_{zz}(z) - \oint \frac{\dd \bar{z}}{2 \pi i}
 \,\, \bar{z} \ep \overline{b}_{\bar{z} \bar{z}} \biggr) \, V[ \tilde{\ep} ] (0) \biggr \rangle 
\nonumber \\
&-
\biggl \langle \biggl( \oint \frac{ \dd z}{2 \pi i} \,\, z  \ep b_{zz}(z) - \oint \frac{\dd \bar{z}}{2 \pi i}
 \,\, \bar{z} \ep \overline{b}_{\bar{z} \bar{z}} \biggr) \, V[ \tilde{\ep} ] (0) \, S_{i} \biggr \rangle.
\label{ope}
\end{align}

\noindent
We will calculate all Feynman diagrams considering the pure spinor action and the $b$-ghost as a power series in the $AdS$
radius. For the parametrization \eqref{coordinates}, the expansion of the action can be found in reference \cite{Mikhailov:2007mr}.
In the above equation $S_i$ represents all contributions of order $1/R$ or greater.

\subsection{Computation.}
The free field propagators can be read from \cite{Mikhailov:2007mr}:
\begin{align}
  \langle x^m(z,\bar{z})x^n(0)\rangle
  & = - \kappa^{mn}\log|z|^2
  \\   
  \langle \theta_L^{\alpha}(z,\bar{z}) \theta_R^{\dot{\beta}}(0)\rangle 
  & = - \kappa^{\alpha\dot{\beta}} \log|z|^2
    \label{ElementaryOPE}
  \\   
  \langle \theta_R^{\dot{\alpha}}(z,\bar{z})
  \theta_L^{\beta}(0)\rangle 
  & = - \kappa^{\dot{\alpha}\beta}\log|z|^2 \,.
\end{align}
The propagator $\lambda w$ can be characterized by saying that for any $A^{\alpha}(\lambda)$ such
that $A^{\alpha}\Gamma^m_{\alpha\beta} \lambda^{\beta}=0$ ({\it i.e.} tangent to the pure spinor cone):
\begin{equation}
     \langle A_{\dot{\alpha}} \left( \lambda(z,\bar{z}) \right) w_+^{\dot{\alpha}}(z,\bar{z})\; \lambda^{\beta} \rangle =
   - \kappa^{\dot{\alpha}\beta}z^{-1}
\end{equation}

\subsubsection{Current Vertex}
Let us focus, for the moment, on contractions that take only one 
$V$ in $V \wedge V$; that is, we are going to compute the OPE
of $(b_{0} - \bar{b}_{0} )$ with $\ep ( \lambda_{3} - \lambda_{1} )$.
The contributions we are interested are represented in the diagrams below:
\begin{figure}[h]
\centering
\includegraphics[scale=1]{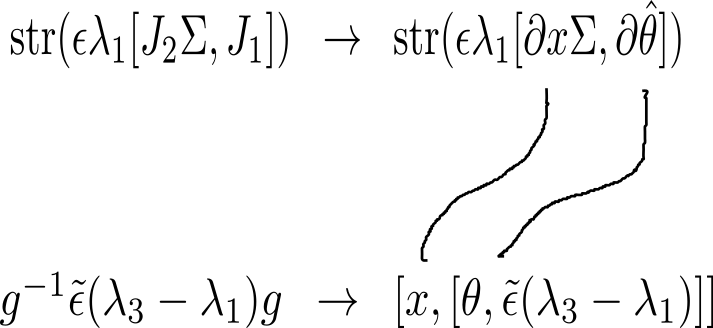}
\caption{\footnotesize{Disconnected contractions for the OPE between $b_{zz}$ and $V[\tilde{\ep}]$.}}
\label{fig:two-matter-contractions}
\end{figure}

\begin{figure}[h]
\centering
\includegraphics[scale=1]{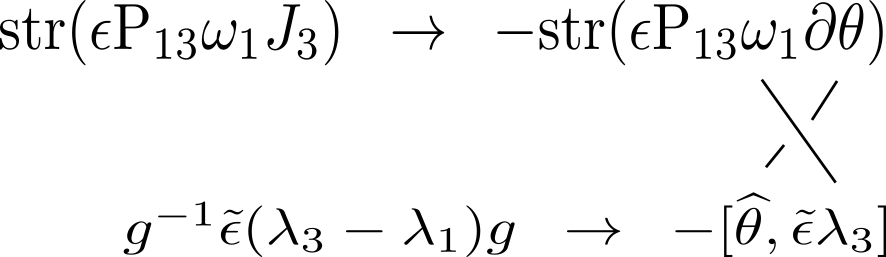}
\caption{\footnotesize{Disconnected contractions for the OPE between $b_{zz}$ and $V[\tilde{\ep}]$.}}
\label{fig:matter-matter-and-ghost-contraction}
\end{figure}

\begin{figure}[h]
\centering
\includegraphics[scale=0.7]{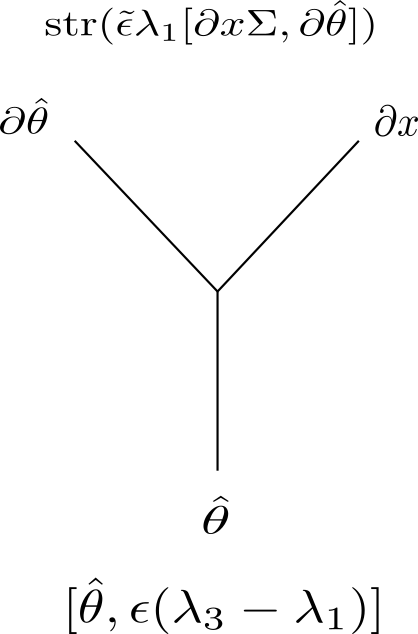}
\caption{\small{Vertex contribution $b_{zz}$ and $V[\tilde{\ep}]$.}}
\label{fig:InternalVertex}
\end{figure}

\paragraph{Contribution from the diagram of Fig. \ref{fig:two-matter-contractions}}
\begin{equation}
- \frac{1}{\mathrm{str} ( \lambda_{3} \lambda_{1})} \frac{\kappa^{mn} \kappa^{\dot{\alpha} \alpha}}{R^{4}(z-w)^{2}} \,
\mathrm{str} \biggl( \ep \lambda_{1} \left[ t^{2}_{m} \Sigma, t^{1}_{\dot{\alpha}} \right] \biggr) 
\biggl[ t^{2}_{n}, \left\{ t^{3}_{\alpha}, \tilde{\ep} \left( \lambda_{3} - \lambda_{1} \right) \right\} \biggr].
\label{firstcontribution}
\end{equation}
Let us use the identity:
\begin{align}
  &-\kappa^{mn} \kappa^{\dot{\alpha} \alpha}\,
    \mathrm{str} \biggl( \ep \lambda_{1} \left[ t^{2}_{m} \Sigma, t^{1}_{\dot{\alpha}} \right] \biggr) 
    \biggl[ t^{2}_{n}, \left\{ t^{3}_{\alpha}, \tilde{\ep} \left( \lambda_{3} - \lambda_{1} \right) \right\} \biggr]  =
  \\   
  =\;
  & \kappa^{mn} \biggl[ t^{2}_{n}, \left[ \left[t_{m}^{2} \Sigma, \ep \lambda_{1} \right] , \tilde{\ep} \left( \lambda_{3} - \lambda_{1} \right) \right] \biggr]  \nonumber
  \\
  =\;
  & \kappa^{mn} \biggl[ t^{2}_{n}, \left[ 
    \left[t_{m}^{2} \Sigma, \ep \lambda_{1} \right] , \tilde{\ep} \lambda_{3}  \right] \biggr].
\end{align}
\paragraph{Contribution from the diagram of Fig. \ref{fig:matter-matter-and-ghost-contraction}}

\begin{align}
    \frac{\kappa^{\alpha \da} \kappa^{\db \beta}}{2 R^{4} (z-\mathrm{w})^{2}}
	\mathrm{str} \left( \ep \mathrm{P}_{13} t^{1}_{\db} t^{3}_{\alpha} \right) \left\{ t^{3}_{\beta}, \tilde{\ep} t^{1}_{\da} \right\}
	=  \frac{\kappa^{\db \beta}}{2 R^{4} (z - \mathrm{w})^{2}} \left[ \ep t^{3}_{\beta}, \tilde{\ep} \mathrm{P}_{13} t^{1}_{\db} \right].
\end{align}

\paragraph{Sum of first and second diagram}

\begin{equation}
	\ep b_{0} V[\tilde{\ep}] = + \frac{1/R^{4}}{\mathrm{str} ( \lambda_{3} \lambda_{1})} 
	\kappa^{mn} \biggl[ t^{2}_{n}, \left[ \left[t_{m}^{2} \Sigma, \ep \lambda_{1} \right] , \tilde{\ep} \lambda_{3}  \right] \biggr] 
	- \frac{\kappa^{\db \beta}}{2 R^{4}} \left[ \tilde{\ep} \mathrm{P}_{13} t^{1}_{\db}, \ep t^{3}_{\beta} \right].
\end{equation}

\paragraph{Anti-holomorphic b-ghost} 
A similar computation gives for the anti-holomorphic term:
\begin{equation}
	\ep \overline{b}_{0} V[\tilde{\ep}] = + \frac{1/R^{4}}{\mathrm{str} ( \lambda_{3} \lambda_{1})} 
	\kappa^{mn} \biggl[ t^{2}_{n}, \left[ \left[t_{m}^{2} \Sigma, \ep \lambda_{3} \right] , \tilde{\ep} \lambda_{1}  \right] \biggr] 
	- \frac{\kappa^{\db \beta}}{2 R^{4}} \left[ \tilde{\ep} \mathrm{P}_{13} t^{1}_{\db}, \ep t^{3}_{\beta} \right]. 
\end{equation}

\paragraph{Contribution of $b_{0} - \overline{b}_{0}$}
We can simplify the total contrubution of the diagrams of Figures \ref{fig:two-matter-contractions} and \ref{fig:matter-matter-and-ghost-contraction} to

\begin{align}
\ep ( b_{0} - \overline{b}_{0} )
V[\te] 
&=
\Bigg(
\kappa^{mn} \Bigg[ t^{2}_{n}, \bigg[ [ t^{2}_{m} \Sigma, \ep \lambda_{1}  ], \te \lambda_{3} \bigg] \Bigg] 
-
\kappa^{mn} \Bigg[ t^{2}_{n}, \bigg[ [ t^{2}_{m} \Sigma, \ep \lambda_{3}  ], \te \lambda_{1}  \bigg] \Bigg] 
\Bigg)
\nonumber \\
\nonumber \\
&=
\Bigg(
\frac{5}{2} \bigg[ \ep \lambda_{1}, \te \lambda_{3} \bigg] -
\frac{5}{2} \bigg[ \ep \lambda_{3}, \te \lambda_{1} \bigg]
\Bigg)
\nonumber \\
\nonumber \\
&+ \Bigg(
\kappa^{mn} \Bigg[ \bigg[ [ t^{2}_{m} \Sigma, \ep \lambda_{1}  \bigg], \bigg[ t^{2}_{n} , \te \lambda_{3} \bigg] \Bigg] 
-
\kappa^{mn} \Bigg[ \bigg[ [ t^{2}_{m} \Sigma, \ep \lambda_{3}  \bigg], \bigg[ t^{2}_{n} , \te \lambda_{1} \bigg] \Bigg] 
\Bigg) 
\nonumber \\
\nonumber \\
&=\Bigg(
\kappa^{mn} \Bigg[ \bigg[ [ t^{2}_{m} \Sigma, \ep \lambda_{1}  \bigg], \bigg[ t^{2}_{n} , \te \lambda_{3} \bigg] \Bigg] 
-
\kappa^{mn} \Bigg[ \bigg[ [ t^{2}_{m} \Sigma, \ep \lambda_{3}  \bigg], \bigg[ t^{2}_{n} , \te \lambda_{1} \bigg] \Bigg] 
\Bigg) 
\nonumber \\
\nonumber \\
&=
- \frac{3}{2} \bigg[ \ep \lambda_{1}, \te \lambda_{3} \bigg]
+\frac{3}{2} \bigg[ \ep \lambda_{3} , \te \lambda_{1} \bigg]
= 0
\end{align}

In this derivation, we used the identities

\begin{subequations}
\begin{align} \label{fivehalves}
	\kappa^{mn} \left[ t_{n}^{2}, \left[ t^{2}_{m} \Sigma , \ep \lambda_{1} \right] \right] &=
	\frac{\Sigma}{2} \kappa^{mn} \left[ \left\{ t_{m}^{2}, t_{n}^{2} \right\}, \ep \lambda_{1} \right] \nonumber \\
	&= \kappa^{mn} \kappa_{mn} \frac{\Sigma}{8} \left[ \Sigma , \ep \lambda_{1} \right] \nonumber \\
	&= \frac{1}{4} \kappa^{mn} \kappa_{mn} \, \ep \lambda_{1} \nonumber \\
        &= \frac{5}{2} \ep \lambda_{1}
\end{align}

\noindent
together with

\begin{equation}
\kappa^{mn} \Bigg[ \bigg[ t_{m}^{2} \Sigma, \ep \lambda_{1} \bigg] , 
\bigg[ t^{2}_{n} , \te \lambda_{3} \bigg] \Bigg] 
=
- \frac{3}{2} \bigg[ \ep \lambda_{1}, \te \lambda_{3} \bigg]
\label{m1}
\end{equation}

\noindent
and

\begin{equation}
- \kappa^{mn} \Bigg[ \bigg[ t^{2}_{m} \Sigma, \ep \lambda_{3} \bigg],
\bigg[ t^{2}_{n}, \te \lambda_{1} \bigg] \Bigg]
=
\frac{3}{2} 
\bigg[ \ep \lambda_{3} , \te \lambda_{1} \bigg]
\label{m2}
\end{equation}
\end{subequations}

\paragraph{Contribution of the diagram of Figure \ref{fig:InternalVertex}}
There only remains the contractions that get contributions from the interaction
vertices:

\begin{equation}
\frac{1}{2\pi R^{4}} \frac{1}{\mathrm{str} ( \lambda_{3} \lambda_{1} )} 
	\mathrm{str} \left( \ep \lambda_{1} \left[ \partial x \Sigma, \partial \bo \right] \right) (z)
	\left[ \bo, \tilde{\ep} \left( \lambda_{3} - \lambda_{1} \right) \right] (\mathrm{w}) \int \dd^{2} u \,\, \mathrm{str} \left( \partial x  \left[ \oo, \bpartial \oo \right] \right)
\end{equation}
We use:
\begin{equation}
   \int \dd^{2} u \, \frac{1}{(z-u)^{3}} \frac{1}{(\bar{\mathrm{w}} - \bar{u})} =
   {\pi\over 2}{1\over (z-\mathrm{w})^{2}}
\end{equation}
and obtain:

\begin{equation}
- \kappa^{mn} \kappa^{\db \beta }\kappa^{\alpha \da} 
\mathrm{str} \left( \ep \lambda_{1} \left[ t^{2}_{m} \Sigma, t_{\db}^{1} \right] \right) 
\big\{ t^{1}_{\da} , \te ( \lambda_{3} - \lambda_{1} ) \big\} 
\mathrm{str} \left( t^{2}_{n} \left\{ t^{3}_{\beta}, t^{3}_{\alpha} \right\} \right).
\end{equation}

\noindent
We temporarily do not write the factor of $1/4R^{2} \mathrm{str}( \lambda_{3} \lambda_{1} )$ 
since it only observes the calculation.
This answer can be rewritten as

\begin{align} \label{acon}
&- \kappa^{mn} \kappa^{\db \beta }\kappa^{\alpha \da} 
\mathrm{str} \left( \ep \lambda_{1} \left[ t^{2}_{m} \Sigma, t_{\db}^{1} \right] \right) 
\big\{ t^{1}_{\da} , \te ( \lambda_{3} - \lambda_{1} ) \big\} 
\mathrm{str} \left( t^{2}_{n} \left\{ t^{3}_{\beta}, t^{3}_{\alpha} \right\} \right) =
\nonumber \\
\nonumber \\
&- \kappa^{mn} \kappa^{\db \beta }\kappa^{\alpha \da} 
\mathrm{str} \left( [ \ep \lambda_{1},  t^{2}_{m} \Sigma] t_{\db}^{1}  \right) 
\big\{ t^{1}_{\da} , \te ( \lambda_{3} - \lambda_{1} ) \big\} 
\mathrm{str} \left( [ t^{2}_{n},  t^{3}_{\beta}] t^{3}_{\alpha} \right) =
\nonumber \\
\nonumber \\
&- \kappa^{mn} \kappa^{\db \beta }
\mathrm{str} \left( [ \ep \lambda_{1},  t^{2}_{m} \Sigma] t_{\db}^{1}  \right) 
\Bigg[ [ t^{2}_{n} , t^{3}_{\beta} ] , \te ( \lambda_{3} - \lambda_{1} )  \Bigg] =
\nonumber \\
\nonumber \\
&-\kappa^{mn} \Bigg[ \left[ t^{2}_{n} , [ \ep \lambda_{1},  t^{2}_{m} \Sigma ] \right], \te ( \lambda_{3} - \lambda_{1} ) \Bigg] =
\nonumber \\
\nonumber \\
&+ \kappa^{mn} \Bigg[ \left[ t^{2}_{n} , [ t_{m}^{2} \Sigma, \ep \lambda_{1}  ]  \right], \te ( \lambda_{3} - \lambda_{1} ) \Bigg] =
\frac{5}{2} \Bigg[ \ep \lambda_{1}, \te ( \lambda_{3} - \lambda_{1} ) \Bigg] =
\nonumber \\
\nonumber \\
&\frac{5}{2} \Big[ \ep \lambda_{1}, \te  \lambda_{3}  \Big] = \frac{5}{2} \Big[ \ep \lambda_{3}, \te  \lambda_{1}  \Big] 
\end{align}

\noindent
to give the contribution -- with all factors restored --

\begin{equation} \label{con3}
\frac{5}{8 R^{4}} \frac{g^{-1} \Big[ \ep \lambda_{3}, \te  \lambda_{1}  \Big]g}{\mathrm{str} ( \lambda_{3} \lambda_{1} ) }.
\end{equation}

\noindent
Notice that in deriving equation
\eqref{acon}
we used identity \eqref{fivehalves}.
To summarize, the contribution of Figure \ref{fig:InternalVertex} is given by equation \eqref{con3}.

\paragraph{Anti-holormophic b-ghost}
One can compute the contribution of $\overline{b}_{\bar{z} \bar{z}} ( \bar{z} )$ in the same way and it gives
\begin{equation}
 	\ep \bar{b}_{0} V[\tilde{\ep}] (\mathrm{w}) =
	\frac{5}{8 R^{4}} \frac{g^{-1} \Big[ \ep \lambda_{3}, \te  \lambda_{1}  \Big]g}{\mathrm{str} ( \lambda_{3} \lambda_{1} ) }
\end{equation}

\paragraph{Final answer}
Combining the three diagrams we arrive at

\begin{align}
	\ep \left( b_{0} - \overline{b}_{0} \right) V[\tilde{\ep}] = 0
\end{align}

\noindent
for the current vertex.

\subsubsection{Beta-deformation Vertex.}
In order to finish the calculation, we only have to compute contractions where the $b$-ghost
hits both $V$ in $V \wedge V$.
These mixed contractions are given by the diagrams below:

\begin{figure}[H]
\centering
\includegraphics[scale=1]{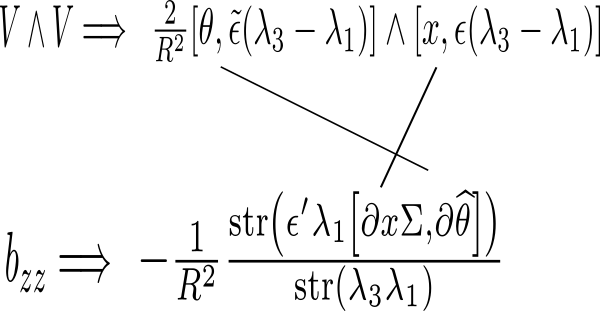}
	\caption{\footnotesize{Disconnected contractions for the OPE between $b_{zz}$ and $V[\tilde{\ep}] \wedge V[\ep]$.}}
\label{fig:contractions-beta-1}
\end{figure}

\begin{figure}[H]
\centering
\includegraphics[scale=1]{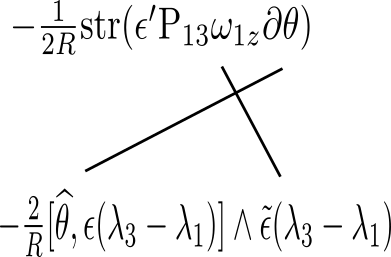}
\caption{\footnotesize{Disconnected contractions for the OPE between $b_{zz}$ and $V[\tilde{\ep}] \wedge V[\ep]$.}}
\label{fig:contractions-beta-2}
\end{figure}

\begin{center}
We stress that there are no contributions from the action up to $1$-loop.
\end{center}

\paragraph{Contribution of diagram in figure \ref{fig:contractions-beta-1}}
The diagram in figure \ref{fig:contractions-beta-1} contributes as

\begin{equation}
- \kappa^{\da \alpha} \kappa^{mn} \frac{2}{R^{4} \mathrm{str} \left( \lambda_{3} \lambda_{1} \right)} \mathrm{str} \left( \ep' \lambda_{1} \left[ t^{2}_{m} \Sigma, t^{1}_{\da} \right] \right)  \,
\{ t^{3}_{\alpha}, \te (\lambda_{3} - \lambda_{1} ) \} \wedge [ t^{2}_{n} , \ep ( \lambda_{3} - \lambda_{1} ) ]  
\end{equation}

\noindent
And this result can be simplified to:

\begin{align}
&-\frac{2\kappa^{\da \alpha} \kappa^{mn}}{R^{4} \mathrm{str} \left( \lambda_{3} \lambda_{1} \right)} \mathrm{str} \left( \ep' \lambda_{1} \left[ t^{2}_{m} \Sigma, t^{1}_{\da} \right] \right) \,
\{ t^{3}_{\alpha}, \te (\lambda_{3} - \lambda_{1} ) \} \wedge [ t^{2}_{n} , \ep ( \lambda_{3} - \lambda_{1} ) ] =
\nonumber \\
\nonumber \\
&-\frac{2\kappa^{\da \alpha} \kappa^{mn}}{R^{4} \mathrm{str} \left( \lambda_{3} \lambda_{1} \right)} \mathrm{str} \left( [ \ep' \lambda_{1}, t^{2}_{m} \Sigma ] t^{1}_{\da} \right) \,
\{ t^{3}_{\alpha}, \te (\lambda_{3} - \lambda_{1} ) \} \wedge [ t^{2}_{n} , \ep ( \lambda_{3} - \lambda_{1} ) ] =
\nonumber \\
\nonumber \\
&-\frac{2 \kappa^{mn}}{R^{4} \mathrm{str} \left( \lambda_{3} \lambda_{1} \right)} 
\bigg[ \big[ \ep' \lambda_{1}, t^{2}_{m} \Sigma \big], \te (\lambda_{3} - \lambda_{1} ) \bigg] \wedge \bigg[ t^{2}_{n} , \ep ( \lambda_{3} - \lambda_{1} ) \bigg]  =
\nonumber \\
\nonumber \\
&-\frac{2 \kappa^{mn}}{R^{4} \mathrm{str} \left( \lambda_{3} \lambda_{1} \right)} 
\bigg[ \big[ \ep' \lambda_{1}, t^{2}_{m} \Sigma \big], \te \lambda_{3} \bigg] \wedge \bigg[ t^{2}_{n} , \ep ( \lambda_{3} - \lambda_{1} ) \bigg] 
\end{align}

\paragraph{Contribution of diagram in figure \ref{fig:contractions-beta-2}}
Likewise, we obtain:

\begin{align}
&\frac{1}{R^{2}} \mathrm{str} \left( \ep' \mathrm{P}_{13} \omega_{1z} \partial \theta \right)
[ \widehat{\theta}, \te ( \lambda_{3} - \lambda_{1} ) ] \wedge \ep ( \lambda_{3} - \lambda_{1} ) =
\nonumber \\
\nonumber \\
&-\frac{1}{R^{2}} \kappa^{\alpha \da} \mathrm{str} \left( \ep' \mathrm{P}_{13} \omega_{1z} t^{3}_{\alpha} \right)
\{ t^{1}_{\da}, \te ( \lambda_{3} - \lambda_{1} ) \} \wedge \ep ( \lambda_{3} - \lambda_{1} ) =
\nonumber \\
\nonumber \\
&\frac{1}{R^{4}} \kappa^{\alpha \da} \kappa^{ \db \beta} \mathrm{str} \left( \ep' \mathrm{P}_{13} t^{1}_{\db} t^{3}_{\alpha} \right)
\{ t^{1}_{\da}, \te ( \lambda_{3} - \lambda_{1} ) \} \wedge \ep  t^{3}_{\beta} =
\nonumber \\
\nonumber \\
&\frac{1}{R^{4}} \kappa^{ \db \beta} 
\bigg[ \ep' \mathrm{P}_{13} t^{1}_{\db} , \te ( \lambda_{3} - \lambda_{1} ) \bigg] \wedge \ep  t^{3}_{\beta} =
\nonumber \\
\nonumber \\
&\frac{1}{R^{4}} \kappa^{ \db \beta} 
\bigg[ \ep' \mathrm{P}_{13} t^{1}_{\db} , \te  \lambda_{3} \bigg] \wedge \ep  t^{3}_{\beta} =
\nonumber \\
\nonumber \\
&\frac{1}{R^{4}} \kappa^{ \db \beta} 
\bigg[ \ep' t^{1}_{\db} , \te  \lambda_{3} \bigg] \wedge \ep  t^{3}_{\beta} 
\end{align}

\paragraph{Holormorphic $b$-ghost} The sum of these contribution gives us:

\begin{align}
\ep' b_{0} V[\te] \wedge V[\ep] &=
-\frac{2 \kappa^{mn}}{R^{4} \mathrm{str} \left( \lambda_{3} \lambda_{1} \right)} 
\bigg[ \big[ \ep' \lambda_{1}, t^{2}_{m} \Sigma \big], \te \lambda_{3} \bigg] \wedge \bigg[ t^{2}_{n} , \ep ( \lambda_{3} - \lambda_{1} ) \bigg] 
\nonumber \\
\nonumber \\
&\,+
\frac{1}{R} \kappa^{ \db \beta} \bigg[ \ep' t^{1}_{\db} , \te  \lambda_{3} \bigg] \wedge \ep  t^{3}_{\beta} 
\end{align}

\paragraph{Anti-holomorphic $b$-ghost}
The same can be done for the anti-holomorphic $b$-ghost, and 
we obtain

\begin{align}
\ep' \bar{b}_{0} V[\te] \wedge V[\ep] &=
- \frac{2 \kappa^{mn} }{R^{4}\mathrm{str} \left( \lambda_{3} \lambda_{1} \right)}  
\bigg[ \left[ \ep' \lambda_{3}, t^{2}_{m} \Sigma \right], \te \lambda_{1} \bigg] \wedge \bigg[ t^{2}_{n} , \ep ( \lambda_{3} - \lambda_{1} ) \bigg] 
\nonumber \\
\nonumber \\
&\,+
\frac{1}{R} \kappa^{\beta \db} \bigg[ \ep'  t^{3}_{\beta},  \te \lambda_{1}  \bigg] \wedge \ep  t^{1}_{\db} 
\end{align}

\subsubsection{Final answer}
The sum of all contributions from the current and the mixed contractions
gives us the final answer:

\begin{align}
&\ep' \left( b_{0} - \overline{b}_{0} \right) V[\te] \wedge V[\ep] = 
\nonumber \\
\nonumber \\
&\frac{-2 \kappa^{mn}}{R^{4} \mathrm{str} \left( \lambda_{3} \lambda_{1} \right)} \Bigg(
\bigg[ \big[ \ep' \lambda_{1}, t^{2}_{m} \Sigma \big], \te \lambda_{3} \bigg] \wedge \bigg[ t^{2}_{n} , \ep ( \lambda_{3} - \lambda_{1} ) \bigg] 
- \bigg[ \left[ \ep' \lambda_{3}, t^{2}_{m} \Sigma \right], \te \lambda_{1} \bigg] \wedge \bigg[ t^{2}_{n} , \ep ( \lambda_{3} - \lambda_{1} ) \bigg]  \Bigg)
\nonumber \\
\nonumber \\
&+ \frac{1}{R^{4}}
\Bigg( \kappa^{ \db \beta} \bigg[ \ep' t^{1}_{\db} , \te  \lambda_{3} \bigg] \wedge \ep  t^{3}_{\beta} 
-
\kappa^{\beta \db} \bigg[ \ep'  t^{3}_{\beta},  \te \lambda_{1}  \bigg] \wedge \ep  t^{1}_{\db} \Bigg)
\end{align}

\section{BRST triviality of $f_{ab}{}^c W_c$}\label{sec:BRST-triviality}
The projectors ${\bf P}$ were used in \cite{Bedoya:2010qz} to prove that BRST triviality of the ghost number $1$ 
vertices corresponding to the global symmetries. 
Once we allow denominators, the BRST cohomology is zero anyway.
But in highly supersymmetric backgrounds, it is meaningful to ask to which extent resolving 
$Q\phi = \psi$ preserves the global supersymmetries. 
The ghost number $1$ vertex for a global symmetry  $t_a\in {\bf psu(2,2|4)}$ is:
\begin{equation}
W_a(\epsilon) = \left(g^{-1}(\epsilon\lambda_3 - \epsilon\lambda_1) g\right)_a
\end{equation}
for a Grassmann odd constant parameter\footnote{As usual in supergeometry, we use a sufficiently
large pool of constant fermionic parameters} $\epsilon$. It was proven in \cite{Bedoya:2010qz} that
\begin{align}
  & {f_{ab}}^c W_c\;=\; 
    - \epsilon Q X_{ab} = - \epsilon Q X_{[ab]}
    \quad\mbox{\tt\small where } 
    \label{QXab}
  \\
  & X_{ab} \;=\; \mbox{Str}\left(
    gt_ag^{-1}\;\left( 
    (gt_bg^{-1})_{\bar{3}} + 2(gt_bg^{-1})_{\bar{2}} + 3(gt_bg^{-1})_{\bar{1}}  
    -4  {\bf P}_{13}(gt_bg^{-1})_{\bar{1}} \right) \right)
    \nonumber 
\end{align}
where $f_{ab}^c$ are the structure constants of ${\bf psu}(2,2|4)$. This implies that $f_{ab}{}^cW_c$  is 
$Q$-exact in a way preserving symmetries. However, $W_c$ cannot be obtained from $f_{ab}{}^cW_c$ preserving
symmetries. (Notice that $f_{ab}{}^cf^{ab}_d = 0$.) In this sense,  $f_{ab}{}^cW_c$ is BRST-exact but $W_c$ is not.

Notice that:
\begin{align}
  X_{[ab]}\;=\;
  & \Big\langle
    t_a\wedge t_b\;,\; g^{-1}Ag
    \Big\rangle
  \label{Xab}\\   
  \mbox{\tt\small where }A
  & = - 2k^{\alpha\dot{\alpha}} t^3_{\alpha}\wedge ({\bf 1}-2{\bf P}_{13})t^1_{\dot{\alpha}} \;=
  \\
  & = 2 k^{\alpha\dot{\alpha}}t^3_{\alpha}\wedge t^1_{\dot{\alpha}}
    + 8 {k^{\alpha\dot{\alpha}} t^3_{\alpha}\wedge
    [\overline{\{\lambda_1,t^1_{\dot{\alpha}}\}}_{\rm STL}\,,\,\lambda_3]
    \over \mbox{STr}\lambda_1\lambda_3}\;=
    \nonumber
  \\
  & = 2 k^{\alpha\dot{\alpha}}t^3_{\alpha}\wedge t^1_{\dot{\alpha}}
    + 8 {[\lambda_1,t_m^2]\wedge [t_{-m}^2,\lambda_3]\over\mbox{STr}\lambda_1\lambda_3}
\end{align}
In other words, 
in the covariant complex (see Section \ref{sec:CovarianceOfVertices}, Eq. (\ref{QOnv})):
\begin{equation}\label{Qtt}
   Q\left(
      k^{\alpha\dot{\alpha}} t^3_{\alpha}\wedge ({\bf 1}-2{\bf P}_{13})t^1_{\dot{\alpha}}
   \right) = d_{\rm Lie} (\lambda_3 - \lambda_1)
\end{equation}
where $d_{\rm Lie}$ is defined in Section \ref{sec:LieAlgebraHomology}. 

\paragraph     {Relation to the ``minimalistic action''}
We will now explain that Eq. (\ref{Qtt}) is equivalent to the BV Master Equation for
the minimalistic action of \cite{Mikhailov:2017mdo}. Let us consider the scalar
product, as defined in Section \ref{sec:LieAlgebraHomology}, with $J_3\wedge J_1$:
\begin{align}
  & \Big\langle
    J_3\wedge J_1\;,\;
    k^{\alpha\dot{\alpha}}t_{\alpha}^3 
    \wedge
    ({\bf 1} - 2 {\bf P}_{13})t_{\dot{\alpha}}^1 
    \Big\rangle\;=\;
  \\
  \;=\;
  & \mbox{STr}\left( J_{1}t_{\alpha}^3 \right)k^{\alpha\dot{\alpha}}\wedge
    \;\mbox{STr}\left( t^1_{\dot{\alpha}} ({\bf 1} - 2{\bf P}_{31}) J_3 \right)\;=\;
  \\
  \;=\;
  & \mbox{STr}\left( J_1\wedge ({\bf 1} - 2{\bf P}_{31}) J_3 \right)
\end{align}
\begin{align}
  & \epsilon Q \Big\langle
    J_3\wedge J_1\;,\;
    k^{\alpha\dot{\alpha}}t_{\alpha}^3 
    \wedge
    ({\bf 1} - 2 {\bf P}_{13})t_{\dot{\alpha}}^1 
    \Big\rangle\;=\;
  \label{PairingWithJ3WedgeJ1}\\
  \;=\;
  & \phantom{+}\Big\langle
    [\epsilon\lambda_1,J_2]\wedge J_1 + J_3\wedge [\epsilon\lambda_3,J_2]\;,\;
    k^{\alpha\dot{\alpha}}t_{\alpha}^3 
    \wedge
    ({\bf 1} - 2 {\bf P}_{13})t_{\dot{\alpha}}^1 
    \Big\rangle\;+
  \nonumber\\   
  & +\;\Big\langle
    - \epsilon D_0\lambda_3\wedge J_1 - J_3\wedge \epsilon D_0\lambda_1 \;,\;
    k^{\alpha\dot{\alpha}}t_{\alpha}^3 
    \wedge
    ({\bf 1} - 2 {\bf P}_{13})t_{\dot{\alpha}}^1 
    \Big\rangle
    \nonumber
\end{align}
The first line of the RHS of Eq. (\ref{PairingWithJ3WedgeJ1}) equals to (in the sense of Section \ref{sec:CovarianceOfVertices}, Eq. (\ref{QOnv})):
\begin{align}
  & - \Big\langle
    J_{2}\wedge J_{1} + J_{3}\wedge J_{2}\;,\;
    \epsilon Q\left(
    k^{\alpha\dot{\alpha}}t_{\alpha}^3
    \wedge
    ({\bf 1} - 2 {\bf P}_{13})t_{\dot{\alpha}}^1 
    \right)
    \Big\rangle\;=\;
  \nonumber\\   
  \;=\;
  & \Big\langle
    J_2\wedge J_1 + J_3\wedge J_2\;,\;
    d_{\rm Lie} (\epsilon \lambda_3 - \epsilon \lambda_1)
    \Big\rangle\;=\;
  \\   
  \;=\;
  & \Big\langle
    [J_2, J_1] + [J_3, J_2]\;,\;\epsilon \lambda_3 - \epsilon \lambda_1
    \Big\rangle\;=\;
    \mbox{STr}\left([J_3, J_2]\epsilon \lambda_3 - [J_2, J_1]\epsilon \lambda_1\right)
    \nonumber
\end{align}
The second line of the RHS of Eq. (\ref{PairingWithJ3WedgeJ1}) is:
\begin{align}
  & \Big\langle
    - \epsilon D_0\lambda_3\wedge J_1 - J_3\wedge \epsilon D_0\lambda_1 \;,\;
    k^{\alpha\dot{\alpha}}t_{\alpha}^3 
    \wedge
    ({\bf 1} - 2 {\bf P}_{13})t_{\dot{\alpha}}^1 
    \Big\rangle \; =
  \\
  = \;
  & \Big\langle
    - \epsilon D_0\lambda_3\wedge J_1 - J_3\wedge \epsilon D_0\lambda_1 \;,\;
    k^{\alpha\dot{\alpha}}t_{\alpha}^3 \wedge t_{\dot{\alpha}}^1
    \Big\rangle\;=
    \nonumber
  \\
  =\;
  & \mbox{STr}\left((D_0\lambda_1)J_3 - (D_0\lambda_3)J_1\right)
\end{align}
The sum is a total derivative:
\begin{equation}
   Q\mbox{STr}\left( J_1\wedge ({\bf 1} - 2{\bf P}_{31}) J_3 \right)\;=\;
   d\mbox{STr}\left((\lambda_3-\lambda_1)J\right)
\end{equation}
This shows that Eq. (\ref{SBV}) is $Q$-invariant.


\def\cprime{$'$} \def\cprime{$'$}
\providecommand{\href}[2]{#2}\begingroup\raggedright\endgroup

\end{document}